\documentclass{jfm}
\usepackage{multirow} 
\usepackage{booktabs}
\usepackage{rotating}
\usepackage{graphicx}
\usepackage[caption=false]{subfig}
\usepackage{newtxtext}
\usepackage{newtxmath}
\usepackage{ragged2e}
\usepackage{blindtext}
\usepackage{natbib}
\usepackage{xcolor}
\usepackage{soul}
\usepackage{hyperref}
\hypersetup{
colorlinks = true,
urlcolor   = blue,
citecolor  = blue,
}

\DeclareUnicodeCharacter{2009}{\,} 
\usepackage[format=plain]{caption}
\makeatletter
\def\pagelimitfooter{}
\makeatother

\title{Confinement-induced evolution and breakup of viscoelastic filaments in microfluidic coflows}
\author{
U. K. Kar\aff{1},
T. Sujith\aff{1},
D. Ghosh\aff{1}
\and
A. K. Sen\aff{1}
\corresp{\email{ashis@iitm.ac.in}}
}

\affiliation{
\aff{1} Micro Nano Bio Fluidics Unit, Department of Mechanical Engineering, Indian Institute of Technology Madras, Chennai-600036, India
}

\begin{document}
\maketitle

\thispagestyle{empty}

\begin{abstract}

Viscoelastic filament thinning is classically described by elastocapillary dynamics in extensional flows, yet in confined microchannels the combined effects of wall-induced shear, elasticity, and capillarity remain poorly understood. Here, we experimentally investigate the breakup of a shear-thinning viscoelastic liquid coflowing with an immiscible Newtonian fluid in a rectangular microchannel, with particular emphasis on the formation, stretching, instability, and breakup of the thin filament connecting the primary droplet to the upstream liquid. Four flow regimes: stable coflow, squeezing, dripping, and jetting, are identified and mapped using the capillary numbers of the dispersed and continuous phases together with an elastocapillary parameter. While elasticity has only a weak influence on the onset of primary droplet formation, it profoundly modifies the subsequent filament dynamics by delaying capillary breakup and stabilising long-lived filaments. Scaling analyses based on capillary, viscous, and elastic force balances provide predictive relations for the primary droplet size, the critical filament thickness at the onset of instability, the maximum filament length, and the critical jet length. Particle-tracking measurements reveal that confinement generates a non-uniform wall-induced shear field along the inclined filament, producing spatial variations in interfacial velocity and causing the first bead-on-a-string instability to develop at the location of maximum shear. Furthermore, a Rayleigh–Plateau analysis incorporating an effective viscosity derived from the Oldroyd–B constitutive model predicts the instability wavelength and growth rate to the correct order of magnitude. Together, these results show that confined viscoelastic breakup is governed not solely by classical elastocapillary thinning but by a coupled wall-shear–elasticity mechanism that controls filament stretching, instability, and secondary droplet formation, thereby providing a predictive framework for filament-mediated breakup in confined viscoelastic multiphase flows. 

\end{abstract}

\begin{keywords}
 Microfluidics, Viscoelastic, Coflow, Bead-on-a-string, Secondary drop
\end{keywords}


\section{Introduction}
\label{sec:1}

Droplet formation in multiphase microflows is a canonical fluid-mechanical problem involving nonlinear interactions among viscous, capillary, inertial, and interfacial forces. Advances in microfabrication and high-speed imaging have transformed droplet microfluidics into a powerful platform for investigating interfacial transport phenomena and generating highly controlled emulsions. Precise manipulation of droplet size and generation frequency underpins a broad range of applications including inkjet printing, spray technologies, chemical synthesis, and biomedical systems \citep{HoathLinks2009,MorrisonInkjet2010,ChenTrends2022}. At microscale dimensions, the large surface-to-volume ratio causes interfacial forces to dominate bulk effects, making droplet generation highly sensitive to fluid properties such as viscosity, interfacial tension, wettability, and channel geometry \citep{KumarPathak2022}. Understanding how these competing mechanisms govern interfacial deformation and breakup therefore remains a fundamental challenge in fluid mechanics.

For Newtonian fluids, droplet generation in microchannels is relatively well understood. The breakup process is governed primarily by the balance among viscous stresses, capillary pressure, and inertia, giving rise to distinct flow regimes such as squeezing, dripping, and jetting \citep{GarsteckiStone2006,CubudTMason2008}. These regimes have been systematically characterized across diverse geometries including T-junctions \citep{Nisisako2002,GarsteckiStone2006,vanSteijnKreutzer2010}, flow-focusing devices \citep{Utada2007}, hydrodynamic focusing systems \citep{CubudTMason2008}, and coflow configurations. During breakup, the interface develops a thinning neck whose evolution approaches finite-time singular behaviour driven by capillary stresses. While this framework successfully describes Newtonian systems, its applicability to complex fluids remains uncertain.

Many fluids of practical importance are viscoelastic and often exhibit pronounced shear-thinning behaviour. In such fluids, deformation generates elastic stresses arising from polymer stretching, while the apparent viscosity simultaneously depends on the local shear rate. Consequently, the force balance governing interfacial deformation becomes considerably more complex, involving coupled effects of viscosity, elasticity, and nonlinear rheology \citep{Steinhaus2007}. The resulting dynamics depend not only on stress magnitudes but also on the competition among characteristic time scales associated with capillary relaxation, polymer relaxation, and imposed flow \citep{Steinhaus2007}. Although increasing attention has been devoted to microflows of complex fluids \citep{WangDong2024,Neshat2025}, it remains unclear whether viscoelastic droplet generation follows breakup scenarios analogous to Newtonian systems or whether elasticity introduces fundamentally different mechanisms and flow regimes.

A distinguishing feature of viscoelastic breakup is the emergence of long-lived liquid filaments connecting droplets prior to rupture. Extensive studies on liquid bridge thinning and capillary breakup extensional rheometry have shown that these filaments undergo prolonged thinning before destabilization \citep{EntovHinch1997,Anna2001,OliveiraMcKinley2005,Tirtaatmadja2006}. Unlike Newtonian fluids, where capillary pressure rapidly drives pinch-off, polymer stretching generates tensile elastic stresses that resist deformation and substantially delays breakup. As a result, filament evolution typically exhibits two successive regimes. During the initial stage, thinning is governed by a viscocapillary balance, resulting in power-law decay of filament radius \citep{Tirtaatmadja2006}. As thinning progresses, elastic stresses increasingly oppose capillary forces, producing a nearly constant extensional strain rate and an exponential decay of filament thickness characteristic of the elastocapillary regime \citep{EntovHinch1997,Amarouchene2001,Tirtaatmadja2006}. These dynamics form the basis of extensional rheometry and provide a direct route for estimating polymer relaxation times.

Subsequent evolution of highly stretched filaments introduces additional complexity. As thinning continues, the filament becomes unstable and develops characteristic beads-on-a-string (BOAS) structures consisting of large beads connected by slender threads \citep{WagnerEggers2005,ClasenEggers2006,ArdekaniSharma2010}. Despite extensive investigation, the physical origin of these instabilities remains debated. Proposed mechanisms involve competing influences of capillary, elastic, inertial, and viscous effects \citep{OliveiraMcKinley2005,ArdekaniSharma2010}. More recently, Deblais \textit{et al.} \citep{DeblaisHerrada2020} suggested that BOAS formation occurs when the inverse relaxation time exceeds the growth rate associated with capillary instability. Nevertheless, whether these structures represent modified Rayleigh–Plateau instabilities or arise through distinct viscoelastic mechanisms remains unresolved \citep{WagnerEggers2005,ChenAshgriz2025}. Linear stability analyses have further shown that instability growth rates depend strongly on both viscous and elastic effects characterized through the Ohnesorge and Deborah numbers \citep{LiYin2017}. Although such studies provide valuable insight into extensional filament dynamics, they primarily concern idealized geometries and freely suspended liquid bridges.

By contrast, viscoelastic filament evolution under confinement remains comparatively unexplored. In confined microchannels, filament thinning occurs in the simultaneous presence of interfacial shear, geometric confinement, and wall-induced velocity gradients, introducing mechanisms absent from extensional rheometry. Steinhaus \textit{et al.} \citeyearpar{Steinhaus2007} demonstrated that confinement can alter the transition between elastocapillary and inertia-capillary thinning mechanisms. Similarly, Arratia \textit{et al.} \citeyearpar{Arratia2008} showed that polymeric filaments in microchannels retain exponential thinning behaviour even under imposed flow. However, previous studies primarily focused on rheological characterization and filament thinning rates, without examining the complete sequence of droplet formation, filament evolution, and secondary breakup.

Recent studies of viscoelastic droplet generation in T-junctions and flow-focusing geometries \citep{VGiri2024,WangDong2024,Neshat2025} established empirical relations for primary droplet size and filament extension. However, predictive scaling frameworks remain limited, and the governing role of confinement-induced shear during filament evolution remains poorly understood. Furthermore, despite the importance of secondary droplets in determining emulsion polydispersity, the mechanisms governing filament destabilization, instability wavelength selection, and secondary droplet formation in confined viscoelastic systems have not been systematically investigated. Coflow geometries are particularly attractive in this context because they provide a controlled environment in which shear and confinement act continuously throughout the breakup process. Yet, compared with T-junction and flow-focusing configurations, viscoelastic coflows have received relatively little attention.

The present work addresses these unresolved questions through a systematic experimental investigation of droplet formation and filament breakup in confined viscoelastic co-flows. Using multiple pairs of immiscible fluids comprising a Newtonian continuous phase and viscoelastic dispersed phases, we systematically vary the polymer relaxation time while maintaining nearly constant zero-shear viscosity and interfacial tension, thereby isolating the role of elasticity. Distinct flow regimes are identified over a wide range of operating conditions, and predictive scaling relations are developed for the primary droplet diameter, critical filament thickness, maximum filament length, and jetting length. Beyond regime characterization, we demonstrate that wall-induced shear actively drives filament thinning and controls the onset of secondary instability. To rationalize these observations, we develop a scaling framework for elastocapillary stretching under shear-dominated confinement, relating the extensional strain rate to the Reynolds, Capillary, and Weissenberg numbers through a physically motivated interfacial shear scaling. Direct particle-tracking measurements provide evidence of the axial flow within thinning filaments, elucidating the coupled roles of capillary pressure, elasticity, and wall-induced shear in filament destabilization and secondary droplet formation. Finally, we characterize the statistics of secondary droplets and demonstrate strategies for reducing droplet polydispersity through appropriate selection of fluid properties and flow conditions. Collectively, these results establish a predictive and mechanistic framework for viscoelastic filament breakup in confined co-flows, revealing the previously unexplored role of wall-induced shear in governing filament instability and secondary droplet generation. 

\section{Experiments}
\label{sec:2}

\begin{figure}
    \centering
    \subfloat{
        \includegraphics[width=0.505\columnwidth]{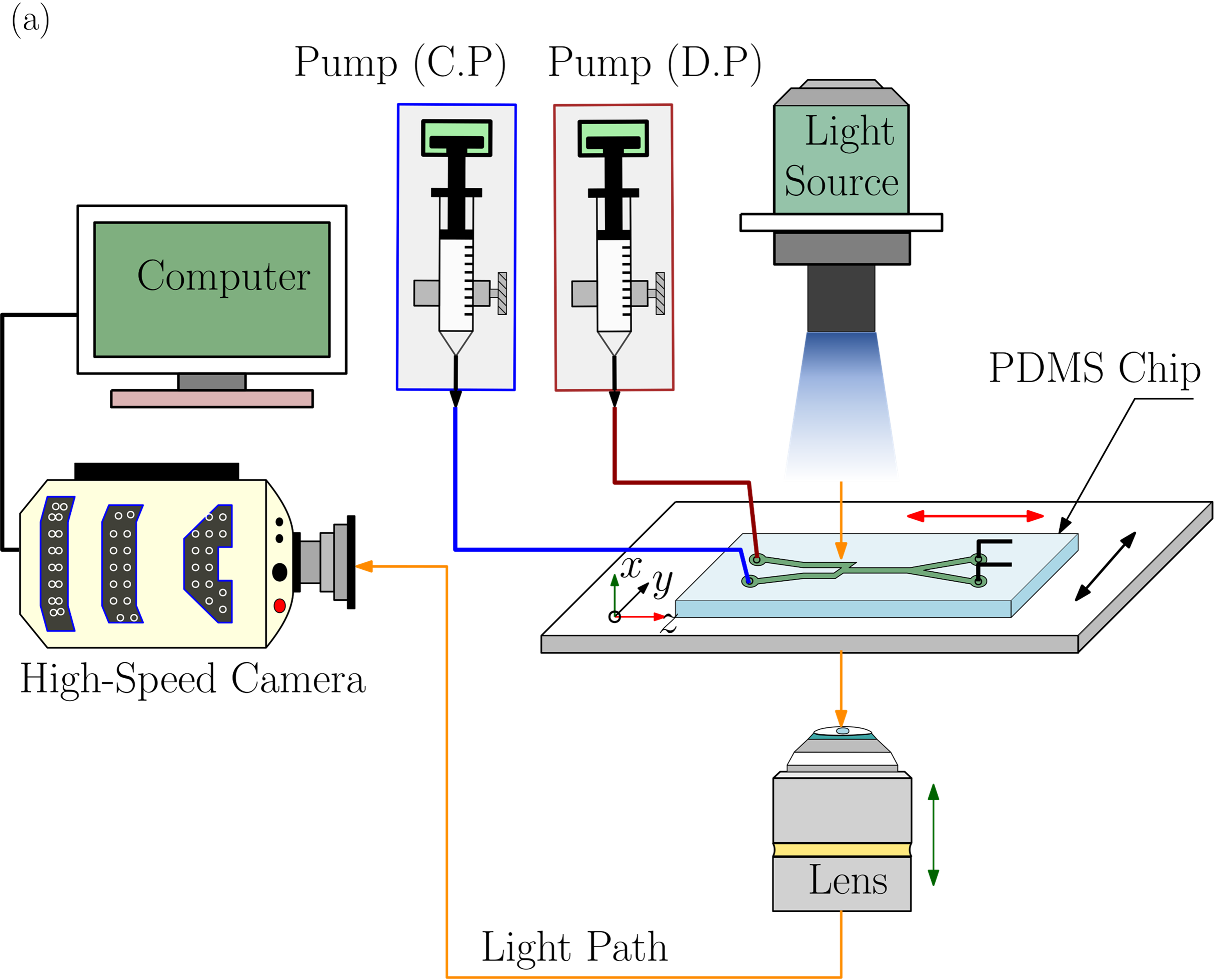}%
        \label{fig:1a}%
    }
    \hfill
    \subfloat{
        \includegraphics[width=0.48\columnwidth]{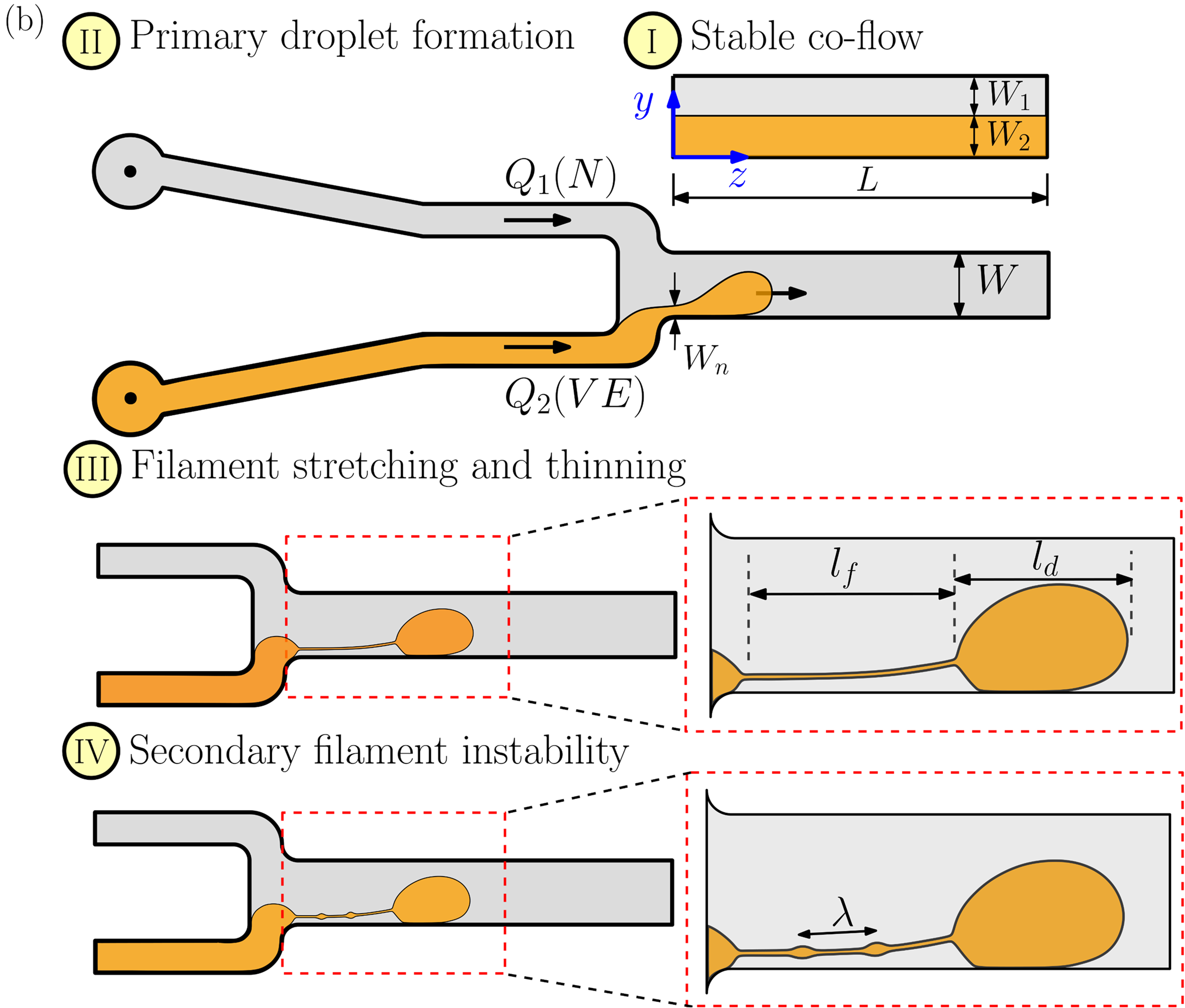}%
        \label{fig:1b}%
    }
    \medskip
    \caption{\justifying{Schematic of the experimental setup and the flow regimes. (a) Schematic of the experimental setup comprising an optical microscope, syringe pumps, a high-speed camera, a PDMS microfluidic chip, and a computer-based data acquisition system. (b) Flow regimes observed in the experiments. (I) Stable coflow regime, where the width ratio of the viscoelastic and Newtonian phases ($W_2/W_1$) remains constant along the main channel of length $L = 27$ mm and width $W = 300~\mu$m. (II--IV) Evolution of droplet formation and filament dynamics: (II) The viscoelastic (VE) stream deforms into a primary droplet connected to the upstream fluid by a neck of width $W_n$. (III) The neck undergoes progressive thinning, resulting in the formation of a stable viscoelastic filament of length $l_f$, while $l_d$ denotes the length of the primary droplet. (IV) The filament subsequently becomes unstable due to capillary-driven instability, giving rise to a bead-on-a-string structure. The distance between two successive beads, denoted by $\lambda$, represents the wavelength of the instability.}}
    \label{fig:1}
\end{figure}

The dynamics of droplet formation and filament breakup were investigated experimentally using a microfluidic co-flow device in which a viscoelastic dispersed phase co-flowed with an immiscible Newtonian continuous phase. A schematic of the experimental setup is shown in Figure \ref{fig:1a}. The device consists of two inlet channels that merge into a straight expansion channel, which serves as the observation region for droplet formation and filament evolution. The observation channel has a length of $L=27~mm$, a width of $W=300~\mu m$, and a depth of $h=100~\mu m$. The microfluidic device was fabricated from polydimethylsiloxane (PDMS) using standard soft-lithography techniques. After plasma-assisted bonding to a PDMS-coated glass substrate, the channel walls were rendered hydrophobic using Aquapel treatment to ensure stable oil-continuous operation. Details of the fabrication procedure are provided in Appendix \ref{appA}.  The dispersed phase comprised aqueous polymer solutions prepared from polyethylene oxide (PEO; molecular weights of 0.4, 1, 2 and 4 MDa) and polyvinylpyrrolidone (PVP; molecular weight 36 kDa), while the continuous phase consisted of medium-chain triglyceride (MCT) oil containing 4 wt.\% Span 85 surfactant. By varying the molecular weight and concentration of PEO in the range $0.97-2.5 ~\text{wt.}\%$ while maintaining nearly constant zero-shear viscosity and interfacial tension, fluids with different relaxation times were obtained, enabling the effects of elasticity to be isolated systematically. The rheological properties of all polymer solutions were characterised using a rotational rheometer (Anton Paar MCR 72), and the interfacial tension between the aqueous and oil phases was measured using a pendant-drop tensiometer (Krüss DSA25). The measured fluid properties are summarised in Table \ref{tab:table1}, while the variations in viscosities with shear rates are presented in Appendix \ref{appB}. Each measurement was repeated at least three times, resulting in uncertainties below $\pm 3\%$. The dispersed and continuous phases were independently supplied using precision syringe pumps (Cetoni GmbH, Germany) to establish co-flow over a wide range of flow-rate ratios. The evolution of the liquid interface, including primary droplet formation, filament thinning, and secondary breakup, was visualised using an inverted microscope (Olympus IX73) coupled to a high-speed camera (Phantom V2640) operating at $3000-8000$ frames $s^{-1}$. Image sequences were analysed to determine the droplet diameter, filament diameter, filament length, instability wavelength, and secondary droplet size distribution.

\begin{table}
  \begin{center}
\def~{\hphantom{0}}
  \begin{tabular}{lccccccccc}
      $Fluids$  & $\rho $   & $\gamma$ &  $\mu_{0}$ & $\beta$  & $\mu_{\infty}$  & $k$ & $a_1$& $a_2$ & $\lambda_r$ \\
       & $[\text{kg/m}^3]$   & $ [\text{mN/m}]$ &  $[\text{mPa-s}]$ &  & $[\text{mPa-s}]$  & $[\text{s}]$ & &  & $[\text{s}]$ \\[5pt]
\hline       
      \textbf{F1}: PVP 0.36 MDa 8.5 $\%$ & 1089 & 30.09 & 77 & 0.29 & 40 &  0.0022& 2.1 & 0.55 & 0.0013\\
      \textbf{F2}: PEO 0.4 MDa 2.5 $\%$ & 1021 & 6.56 & 73 & 0.29 & 28 & 0.0028 & 2.0 & 0.45 & 0.0044\\
      \textbf{F3}: PEO 1.0 MDa 1.7 $\%$ & 1008 & 7.25 & 75 & 0.29 & 18 & 0.0085& 1.55& 0.40 & 0.0100 \\
      \textbf{F4}: PEO 2.0 MDa 1.15 $\%$ & 1007 & 4.77 & 78 & 0.29 & 16  & 0.0095 & 1.5 & 0.25& 0.0225\\
      \textbf{F5}: PEO 4.0 MDa 0.97 $\%$ & 1005 & 4.53 & 77 & 0.29 & 12  &0.0125 & 1.4 & 0.25 & 0.0540\\
      \textbf{F6}: MCT oil with 4$\%$ Span 85 & 947 & - & 22 & 1 & - & -& & -& -\\
 \hline
  \end{tabular}
  \caption{\justifying{Physical properties of the fluids used in the different experiments. Here, $\rho$, $\mu$, $\gamma$, $\lambda_r$, and $\beta$ denote the density, viscosity, interfacial tension (IFT) with respect to MCT oil, relaxation time, and viscosity ratio, respectively. The symbols $\mu_0$ and $\mu_{\infty}$ represent the zero-shear and infinite-shear viscosities, respectively. The parameters $k_1$, $a_1$, and $a_2$ correspond to the Carreau--Yasuda model \citep{ChenAshgriz2025} used to describe the shear-thinning behaviour of the viscoelastic phase (P2). Here, $k_1$ is the characteristic time constant marking the onset of the shear-thinning regime, $a_2$ is the power-law index describing the degree of shear thinning, and $a_1$ is the Yasuda parameter governing the transition between the Newtonian plateau and the shear-thinning region.}}
  \label{tab:table1}
  \end{center}
\end{table}

\section{Results and Discussion}
\label{sec:3}

We investigate the breakup of a viscoelastic dispersed phase flowing alongside an immiscible Newtonian continuous phase in a confined co-flow microchannel. Throughout this study, MCT oil containing 4 wt\% Span 85 serves as the Newtonian continuous phase (P1), while aqueous solutions of polyethylene oxide (PEO) or polyvinylpyrrolidone (PVP) constitute the viscoelastic dispersed phase (P2). The continuous- and dispersed-phase flow rates are varied over the ranges $Q_1=0.5-60~ \mathrm{\mu l}\, \mathrm{min}^{-1}$ and $Q_2=0.5-60~ \mathrm{\mu l}\, \mathrm{min}^{-1}$, respectively, spanning conditions from stable parallel co-flow to jetting accompanied by extensive filament breakup. 

A central objective of the present study is to isolate the influence of elasticity on droplet breakup. To this end, fluids with different polymer relaxation times are employed while maintaining nearly constant zero-shear viscosity and interfacial tension (see Table \ref{tab:table1}). The dispersed-phase viscosity is characterized over a wide range of shear rates using rotational rheometry. As shown in Appendix \ref{appB}, all fluids exhibit a Newtonian plateau for $\dot{\gamma}\leq30\ \mathrm{s}^{-1}$, whereas shear-thinning becomes increasingly important at higher shear rates. Since the experiments span both regimes, the viscosity is described using the Carreau–Yasuda constitutive relation  \citep{ChenAshgriz2025},
\begin{equation}
\mu_2  =  \mu_{\infty} + \left(\mu_0 - \mu_{\infty} \right) \left(1+(k_1\dot{\gamma})^{a_1}\right)^{(a_2 -1)/a_1}
\label{eqn:3.1}   
\end{equation}
where $\mu_0$ and $\mu_{\infty}$ denote the zero- and infinite-shear viscosities, respectively, $k_1$ is the characteristic transition time, $a_1$ controls the breadth of the transition, and $a_2$ specifies the power-law behaviour.  The fitted parameters are listed in Table \ref{tab:table1} and are used throughout the subsequent scaling analyses to account for the shear-dependent rheology.

Under the present operating conditions, four distinct flow regimes are observed. At low flow rates of both phases, the interface remains stable and the two fluids co-flow in parallel without breakup (Figure \ref{fig:1b}-I). Increasing the continuous-phase flow progressively narrows the viscoelastic stream until the interface becomes unstable and periodic droplet generation is initiated. During pinch-off, the emerging primary droplet remains connected to the upstream fluid by a slender liquid bridge, which subsequently stretches into a long viscoelastic filament as the droplet is convected downstream. Continued stretching causes the filament to undergo progressive thinning before developing interfacial undulations that eventually amplify and rupture the filament into multiple secondary droplets. The breakup process therefore comprises three successive but strongly coupled stages (Figure \ref{fig:1b}: II-IV): primary droplet formation, filament stretching and thinning, and secondary filament instability. While the first stage resembles droplet formation in Newtonian systems, the latter two are profoundly modified by viscoelasticity and constitute the primary focus of the present study. To characterize these processes quantitatively, the primary droplet size, filament thickness, filament length, instability wavelength, and secondary droplet statistics are measured from high-speed image sequences. These quantities form the basis of the scaling analyses and stability arguments developed in the following sections.

The dynamics are governed by the competition among viscous, capillary, elastic and confinement-induced shear stresses. The continuous- and dispersed-phase Capillary numbers are defined as $Ca_i = \mu_i u_i/\gamma$, where $u_i = Q_i/(Wh)$ is the superficial velocity of phase $i$, $Q_i$ is the flow rate of phase $i$, $\mu_i$ is the fluid viscosity, and $\gamma$ is the interfacial tension between the two immiscible phases \citep{CubudTMason2008}. To account for shear-thinning, a modified dispersed-phase Capillary number, $Ca_{2,\dot{\gamma}} = \mu_{2,\dot{\gamma}} u_2/\gamma$, is employed throughout the analysis. Elastic effects are quantified using the Weissenberg number, $Wi = \lambda_r \dot{\gamma}$, which compares the polymer relaxation time with the characteristic deformation time \citep{McKinley2005}. Here, $\lambda_r$ is the polymer relaxation time and $\dot{\gamma}$ is the shear rate. Elastocapillary number \citep{McKinley2005,Steinhaus2007}, $Ec = \lambda_r/t_{\tau} = \lambda_r \gamma/\mu_0 W$, measures the relative importance of elastic and capillary time scales ($t_{\tau}$). The remaining control parameters include the flow-rate ratio $\varphi = Q_1/Q_2$ and the viscosity ratio $\beta = \mu_1/\mu_2$. The corresponding parameter ranges investigated in this work are summarized in  Table \ref{tab:table2}.

The results are organized according to the underlying physics. $\S$ \ref{sec:3-1} establishes the flow-regime map and identifies the transitions between stable co-flow, squeezing, dripping and jetting. Sections $\S$ \ref{sec:3-2} and $\S$ \ref{sec:3-3} develop predictive scaling relations for the primary droplet diameter, critical filament thickness, maximum filament length and jetting length. Section $\S$ \ref{sec:3-4} examines the evolution of the axial velocity field within the thinning filament to elucidate the mechanisms governing filament instability and bead formation. Finally, $\S$ \ref{sec:3-5} analyses the statistics of secondary droplet generation and demonstrates how filament dynamics may be exploited to control droplet polydispersity.

 \label{eqn:1}   

\begin{figure}
    \centering
    \subfloat{
        \includegraphics[width=1.0\columnwidth]{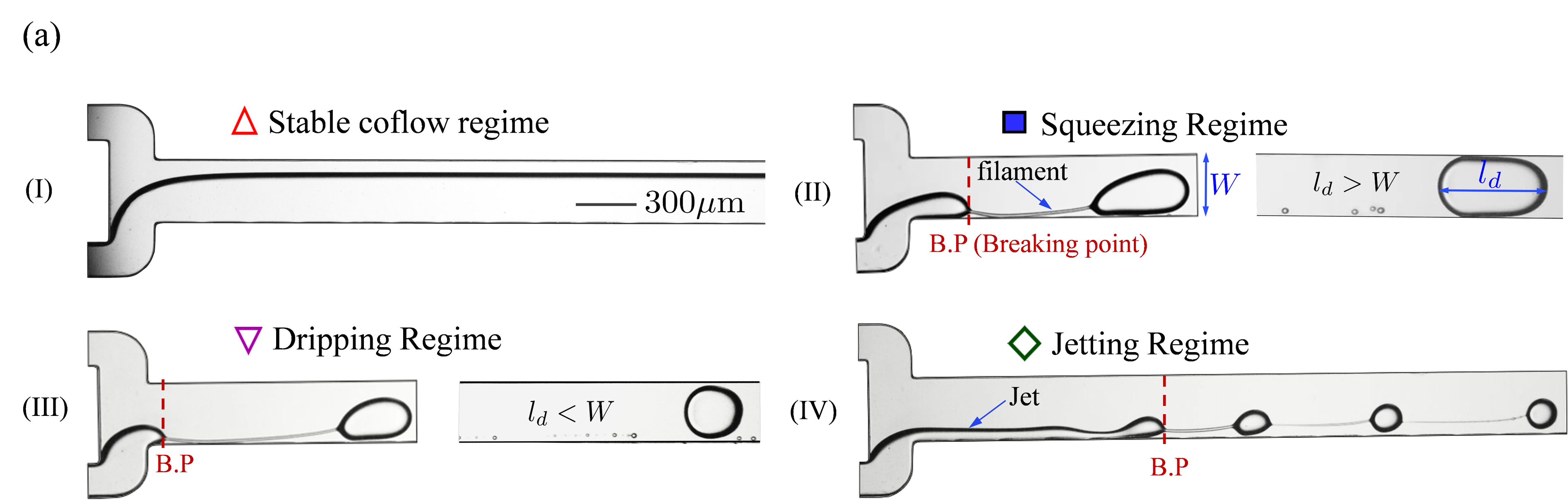}%
        \label{fig:2a}%
    }\\
    
    \subfloat{
        \includegraphics[width=0.85\columnwidth]{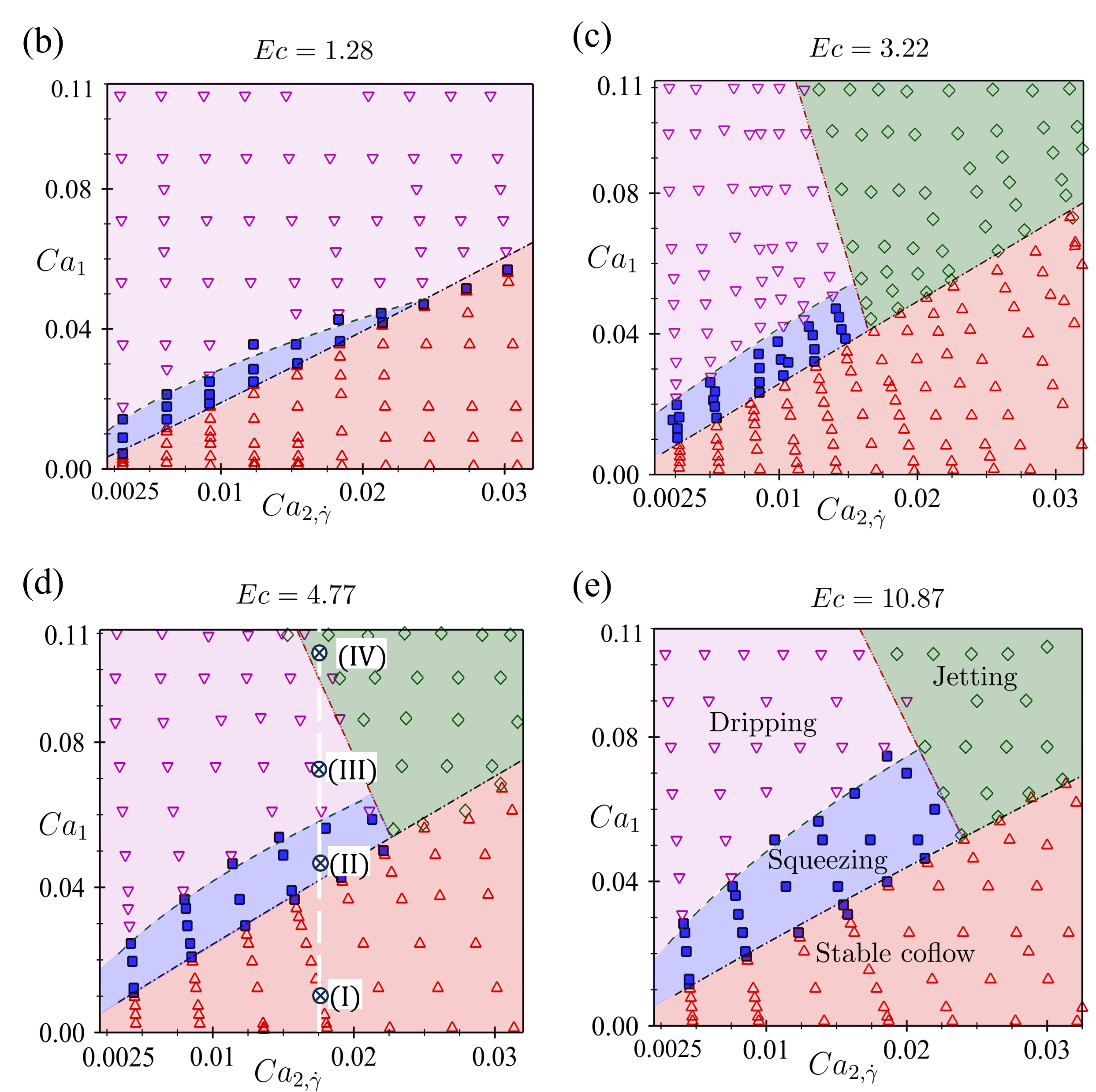}%
        \label{fig:2b}%
    }
    \medskip
    \caption{\justifying{Flow regimes and regime maps based on Capillary and Elastocapillary numbers. (a) Experimental images illustrating the four observed flow regimes at $Ec = 4.77$: stable coflow, squeezing, dripping, and jetting. For a fixed dispersed-phase capillary number, $Ca_2 = 0.0176$, these regimes are observed at $Ca_1 = 0.0012$, $0.0451$, $0.0734$, and $0.1052$, respectively. (b--e) Regime maps in the ($Ca_{2,\dot{\gamma}},,Ca_1$) parameter space for fluids with increasing elasticity ($Ec = 1.28$, $3.22$, $4.77$, and $10.87$), showing the transitions from stable coflow to the droplet-generation regimes.}}
    \label{fig:2}
\end{figure}

\subsection{Flow-regime transitions in confined viscoelastic co-flows}
\label{sec:3-1}

To establish the influence of elasticity on droplet generation, experiments were performed over a wide range of continuous- and dispersed-phase flow rates. Four distinct flow regimes were identified: stable co-flow, squeezing, dripping, and jetting (Figure~\ref{fig:2}(a)). As the continuous-phase flow rate is increased at fixed dispersed-phase flow rate, the system undergoes successive transitions from a stable parallel interface to periodic droplet generation, followed by jet formation at sufficiently large flow rates.

The stable co-flow regime is characterized by a stationary interface that remains uniform along the channel (Figure~\ref{fig:2}(a):I). Beyond a critical forcing, the interface becomes unstable and periodic droplet generation begins. Unlike Newtonian co-flows, where droplet pinch-off occurs shortly after neck formation \citep{CubudTMason2008,Nisisako2002,Christopher2008}, breakup in the present viscoelastic fluids is consistently mediated by the formation of a long, slender filament connecting the primary droplet to the upstream liquid. This filament subsequently undergoes progressive thinning before destabilizing into one or more secondary droplets. Consequently, although the overall flow regimes resemble those observed in Newtonian systems, the breakup mechanism is fundamentally altered by the presence of elasticity. The subsequent sections therefore distinguish between the physics governing primary droplet formation and that governing filament evolution.

The squeezing and dripping regimes (Figure~\ref{fig:2}(a):II and III) are differentiated by the relative size of the generated droplets ($l_d$) with respect to the channel dimensions ($W$). In the squeezing regime, confinement dominates the breakup process, producing elongated plug-like droplets that occupy most of the channel cross-section ($l_d >W$). As the viscous forcing increases, breakup becomes increasingly localized near the junction and nearly spherical droplets ($l_d <W$) characteristic of the dripping regime are produced. At still larger flow rates, the dispersed phase no longer pinches off immediately but instead forms a continuous liquid thread that extends downstream before breaking through capillary instability, defining the jetting regime (Figure~\ref{fig:2}(a):IV). Compared with Newtonian co-flows, the viscoelastic jet remains connected to the upstream fluid by a long-lived filament whose subsequent evolution governs the generation of secondary droplets. The temporal evolution of the observed flow regimes is illustrated in Supplementary Movie S1 for reference.

The observed regimes are summarized in the regime maps of Figure~\ref{fig:2}(b)-(e), expressed in terms of the Capillary numbers of the continuous and dispersed phases. Presenting the transition boundaries in terms of Capillary numbers rather than flow rates highlights the governing competition between viscous and capillary stresses while allowing the influence of viscoelasticity to be examined systematically through the elastocapillary number, $Ec$. The investigated fluids span nearly an order of magnitude in elasticity, from $Ec=1.28$ to $Ec=10.87$. 

A striking feature of the regime maps is that the transition from stable co-flow to droplet generation is largely insensitive to elasticity. For all fluids investigated, the experimentally determined transition boundaries collapse onto a common linear relation, $Ca_1 \approx 2.3~Ca_{2,\dot{\gamma}}$  (see Appendix \ref{appC}), indicating that the onset of interfacial instability is governed primarily by the competition between the viscous forcing imposed by the continuous phase and the capillary resistance of the interface. Since polymer stretching remains weak before significant interface deformation occurs, elastic stresses contribute little to the initiation of primary droplet formation. Instead, elasticity becomes important only after a slender filament has formed, influencing its subsequent evolution rather than the onset of instability, that leads to primary droplet generation.

In contrast, elasticity strongly modifies the transitions within the droplet-generation regime. Increasing $Ec$ progressively delays the transition from squeezing to dripping, shifting the corresponding boundary towards larger values of $Ca_1$. This behaviour reflects the increasing resistance of polymeric stresses to extensional deformation, allowing larger droplets and longer liquid bridges to develop before pinch-off occurs. The influence of elasticity is even more pronounced for the dripping-to-jetting transition. Whereas the least elastic fluid exhibits only squeezing and dripping, a distinct jetting regime emerges at intermediate elasticity before contracting again at the highest elasticity.

The non-monotonic evolution of the jetting regime reflects the coupled effects of elasticity and shear-thinning. At low elasticity, tensile polymer stresses are insufficient to oppose capillary contraction, preventing the formation of a sustained liquid thread. Increasing elasticity stabilizes the thread by delaying capillary necking, thereby promoting jet formation once elastic and capillary stresses become comparable. However, further increases in elasticity are accompanied by stronger shear-thinning, which reduces the effective viscosity in regions of intense deformation and consequently diminishes viscous stabilization of the jet. As a result, progressively larger Capillary numbers are required to maintain a continuous thread, causing the jetting boundary to shift towards higher flow rates. This behaviour is qualitatively consistent with the stability analysis of \citep{MONTANERO2008}, while demonstrating that, for shear-thinning viscoelastic fluids, the extent of the jetting regime is governed not by elasticity alone but by its coupled interaction with shear-dependent viscosity.

The regime maps reveal two distinct roles of viscoelasticity. Elastic stresses exert little influence on the onset of primary droplet formation, which remains controlled primarily by viscous and capillary forces, but become increasingly important during the subsequent evolution of the liquid filament, where they determine the transitions between squeezing, dripping, and jetting. This distinction provides the basis for the scaling analyses of primary droplet formation and filament dynamics developed in the following sections.

\subsection{Dynamics of the squeezing and dripping regimes}
\label{sec:3-2}

The squeezing and dripping regimes comprise two distinct but closely coupled stages of breakup: the formation of the primary droplet at the junction and the subsequent evolution of the viscoelastic filament connecting the droplet to the upstream fluid. Although these processes occur sequentially, they are governed by different physical mechanisms. Primary droplet formation is controlled predominantly by the competition between viscous and capillary stresses, whereas the subsequent filament stretching, thinning, and destabilization are strongly influenced by elastic stresses generated by polymer extension. Accordingly, the droplet and filament are treated separately in the following analysis. We first develop predictive scaling relations for the primary droplet size by considering the dominant force balance during pinch-off. The analysis then focuses on the viscoelastic filament, whose formation distinguishes the present system from Newtonian co-flows, where pinch-off generally occurs without the development of a long-lived connecting thread. Particular emphasis is placed on two quantities that govern the subsequent breakup dynamics: the maximum stable filament length attained before the onset of interfacial instability and the critical filament diameter at which elastic stresses can no longer suppress capillary contraction. The former is estimated from the characteristic stretching time and advection velocity of the filament, whereas the latter is obtained by balancing elastic tensile stresses with capillary pressure. Together, these scaling relations establish a predictive framework linking primary droplet formation to the subsequent evolution and destabilization of confined viscoelastic filaments.

\subsubsection{Primary droplet size scaling }
\label{sec:3-2-1}

Primary droplet formation results from the competition between hydrodynamic forcing exerted by the continuous phase and the restoring effects of surface tension and fluid elasticity. As the dispersed phase enters the main channel, it progressively obstructs the continuous-phase flow, increasing the hydrodynamic resistance and generating a pressure gradient that promotes droplet growth. Simultaneously, viscous shear within the narrow lubrication layer between the interface and the channel wall elongates the interface, whereas capillary pressure and polymer-induced elastic stresses resist deformation. Unlike Newtonian coflows, droplet detachment is invariably followed by the formation of a slender viscoelastic filament. The present section focuses exclusively on predicting the primary droplet size immediately prior to pinch-off, while the subsequent filament dynamics are addressed in the following sections.

\begin{figure}
  \centerline{\includegraphics[width=1\columnwidth]{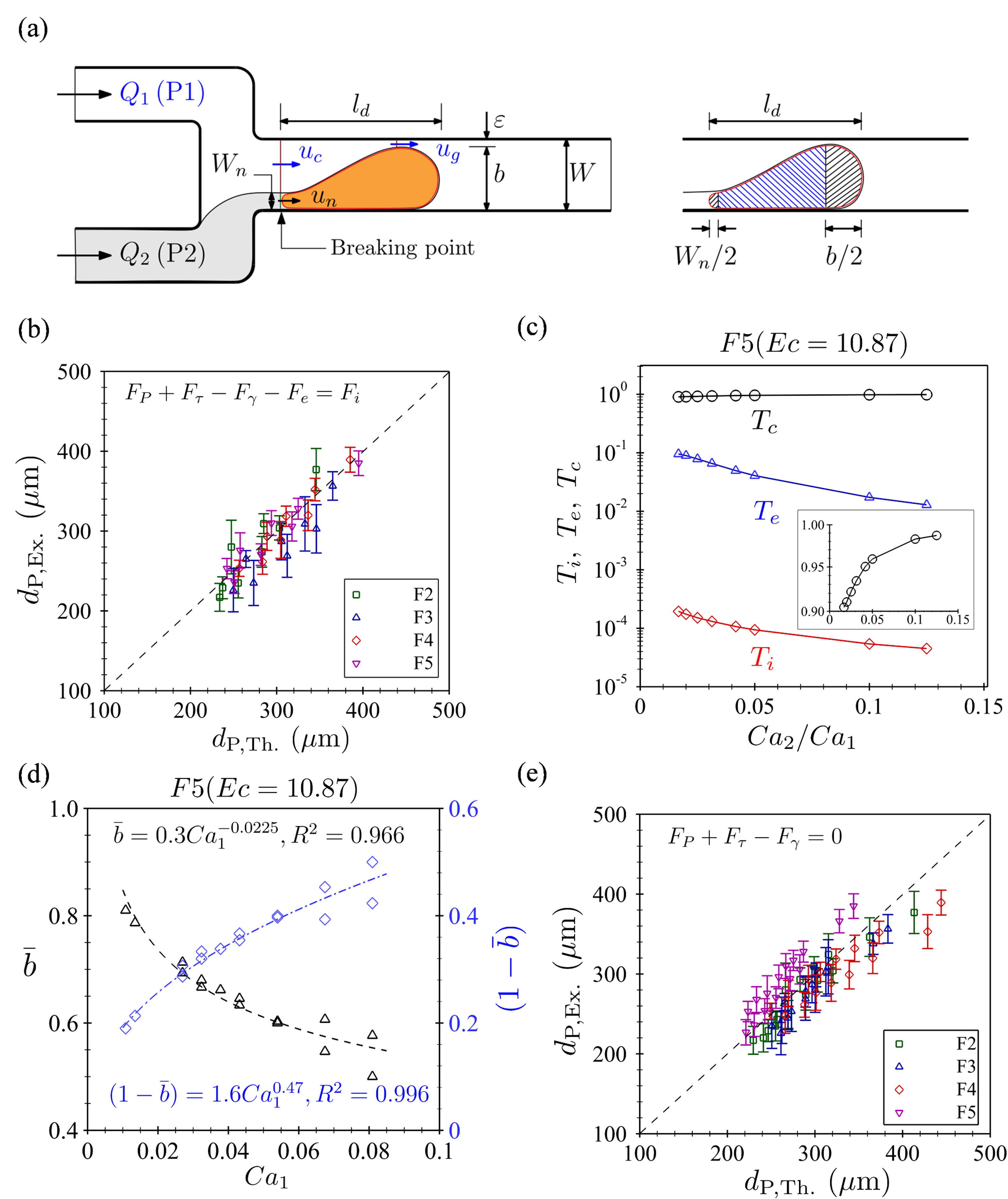}}
   \medskip
  \caption{\justifying{Primary droplet formation in squeezing and dripping regimes: (a) Schematic illustration of primary droplet deformation in a confined microfluidic coflow device immediately before breakup, showing the characteristic length scales, channel geometry, and relevant fluid velocities. (b) Theoretical fit describing the dependence of dimensionless droplet width $\bar{b}$ and gap $\bar{\varepsilon}$ on the continuous-phase Capillary number $Ca_1$. (c) Comparison between the experimentally measured primary droplet diameter, $d_{p,\mathrm{Ex}}$, and the theoretical prediction, $d_{p,\mathrm{Th}}$, obtained from \eqref{eqn:10}. (d) Relative contributions of the inertial, capillary, and elastic terms ($T_i$, $T_c$, and $T_e$, respectively) appearing in \eqref{eqn:10} for $Ec = 10.87$. (e) Comparison between $d_{\text{p,Ex.}}$ and $d_{\text{p,Th.}}$ showing that the theoretical prediction changes only marginally when the inertial and elastic terms are omitted from \eqref{eqn:10}.}}
\label{fig:3}
\end{figure}

A control volume surrounding the growing droplet immediately before detachment is considered (see Figure \ref{fig:3}(a)). Four dominant forces act on the droplet: a pressure force arising from blockage of the continuous-phase flow, viscous shear exerted by the continuous phase, capillary forces resisting deformation, and elastic stresses generated by polymer stretching. The droplet size is determined by the balance among these competing effects.

Following \cite{GarsteckiStone2006}, the pressure force is estimated from the pressure rise produced by partial obstruction of the channel. Unlike previous analyses, however, the present experiments show that the interface is strongly curved only over approximately one-half of the droplet length, while the remaining interface is nearly flat. The characteristic length over which the pressure varies is therefore taken as $l_d/2$ rather than the channel width, yielding

\begin{equation}
F_p \approx \Delta p \, b \,h \approx \frac{\mu_1 Q_1 (l_d/2) }{h^2\varepsilon^2}   bh \approx  \mu_1 u_1 \left(\frac{l_d}{2}\right) \left(\frac{Wb}{\varepsilon^2}\right),
\label{eqn:2}
\end{equation}
where $\Delta p$ is the pressure drop across the droplet, $b$ is the droplet width, and $\varepsilon$ is the minimum lubrication-film thickness between the droplet interface and the channel wall (see Figure~\ref{fig:3}(a)]).

The continuous phase additionally exerts viscous shear on the growing droplet through the thin lubrication layer adjacent to the channel wall. Since the interface simultaneously advances because of the dispersed-phase inflow, the effective shear depends on the relative motion of the two fluids rather than the continuous-phase velocity alone. Particle motion observed near the interface indicates the presence of internal recirculation within the droplet, allowing the interfacial velocity of the dispersed phase to be estimated as $u_{2,i} = 2Q_2/bh$, while the mean velocity within the lubrication layer is approximated as $u_{g} = Q_1 / h\varepsilon$. The corresponding viscous force therefore becomes

\begin{equation}
F_{\tau} \approx 2  \frac{\mu_1 (u_{g}-u_{2,i}) }{\varepsilon} \frac{l_d h}{2} \approx \frac{\mu_1 l_d h}{\varepsilon}\left(\frac{Q_1}{\varepsilon h}-\frac{2Q_2}{bh}\right)
 \approx \mu_1 u_1 l_d \left(\frac{Wh}{\varepsilon^2}-\frac{2u_2}{u_1}\frac{Wh}{b\varepsilon}\right),
 \label{eqn:3}
\end{equation}
The restoring capillary force originates from the Laplace pressure difference between the upstream and downstream interfaces. Assuming identical transverse curvature at both ends of the droplet, the difference in axial curvature is approximately $1/b$, giving

\begin{equation}
F_{\gamma} = \gamma \left(-\frac{1}{b}\right) bh= -\gamma h.
\label{eqn:4}
\end{equation}
which acts opposite to the pressure and viscous forces.

Because the dispersed phase is viscoelastic, polymer stretching generates tensile normal stresses that resist droplet deformation. Following the Oldroyd-B constitutive model and neglecting transverse normal stresses \citep{Aggarwal2008,James2009}, the axial elastic stress is approximated as $\sigma_{zz} \approx 2\mu_p \dot{\gamma_{zz}}^2 \lambda_r$ ,which yields the elastic force

\begin{equation}
F_{e} =  2 \mu_p \dot{\gamma_{zz}}^2 \lambda_r bh,
\label{eqn:5}
\end{equation}
where $\mu_p$ is the polymer contribution to the viscosity of fluid P2, with the total viscosity given by $\mu=\mu_s+\mu_p$, where $\mu_s$ is the solvent contribution to the viscosity.

The characteristic extensional strain rate is estimated from the velocity difference across the growing droplet, $\dot{\gamma}_{zz} \approx  u_r/l_{d} \approx 0.2\,(u_g - u_c)/3b$, where $u_c = Q_1/(W-W_n)h$ and the prefactor accounts for the experimentally observed reduction of the interfacial velocity relative to the bulk flow.

Immediately before pinch-off, the force balance on the growing droplet is obtained from Newton's second law,

\begin{equation}
F_p + F_{\tau} - F_{\gamma} - F_e = \frac{d(mu_d)}{dt},
\label{eqn:7}
\end{equation}
where $m$ denotes the mass of the emerging droplet and $u_d$ its axial velocity. Since the droplet translates with an approximately constant velocity \citep{CubudTMason2008}, the acceleration term is negligible, and the temporal variation of momentum is governed primarily by the increase in droplet mass due to the inflow through the neck, with $dm/dt = \rho_d u_n W_n h$, where $\rho_d$ is the density of the dispersed phase (P2), $W_n$ is the neck width, and $u_n$ is the average velocity of the P2 within the neck.

Substituting the individual force estimates into the momentum equation (\ref{eqn:7}) gives

\begin{equation}
  \mu_1 u_1 \left(\frac{l_d}{2}\right) \left(\frac{Wb}{\varepsilon^2}\right) + \mu_1 u_1 l_d \left(\frac{Wh}{\varepsilon^2}-\frac{2u_2}{u_1}\frac{Wh}{b\varepsilon}\right) -\gamma h -  2 \mu_p \dot{\gamma}^2 \lambda_r bh = \rho_d W_n h u_n u_d
\label{eqn:9} 
\end{equation}

After rearrangement, the droplet length at detachment is obtained as

\begin{equation}
 l_d = \frac{h}{C}\Bigg[ Re_d \left(\frac{\mu_2}{\mu_1}\right) \left(\frac{u_d}{u_1}\right) + \frac{1}{Ca_1} + \frac{2}{3} \left(\frac{\mu_p}{\mu_1}\right) \left(\frac{u_r}{u_1}\right)  Wi\bigg],
\label{eqn:10} 
\end{equation}
where

\begin{equation}
 C = \Bigg[ \frac{Wb}{2\varepsilon^2}+\frac{Wh}{\varepsilon^2}-2\left(\frac{u_2}{u_1}\right) \frac{Wh}{b\varepsilon}\bigg].
\label{eqn:11} 
\end{equation}
The remaining geometric parameters are determined independently. The droplet width is obtained from the modified confinement model \citep{Christopher2008, KumarPathak2022}

\begin{equation}
1.5\, Ca_1 \,\bar{b} = (1-\bar{b})^3,
\label{eqn:6} 
\end{equation}
where $\bar{b} = b/W$, while the neck width is estimated from the experimentally validated correlation $W_n/W = (1 + 3.6(\varphi \beta)^{0.4})^{-1}$, which is shown in Appendix \ref{appD}, to be essentially independent of fluid elasticity. Combining these relations with the geometric approximation illustrated in Figure~\ref{fig:3}(a) yields the theoretical equivalent droplet diameter.

The comparison between the theoretical prediction and experimental measurements in Figure~\ref{fig:3}(b) demonstrates excellent agreement over the entire range of operating conditions, indicating that the proposed force balance successfully captures the dominant physics governing droplet formation. To quantify the relative importance of the competing mechanisms, the individual contributions of inertia ($T_i$), capillarity ($T_c$), and elasticity ($T_e$) are extracted from the governing equation \eqref{eqn:10} and shown in Figure~\ref{fig:3}(c). The first, second, and third terms on the right-hand side of \eqref{eqn:10} represent $T_i$, $T_c$, and $T_e$, respectively. Capillary forces remain the dominant resistance throughout the investigated parameter range, whereas inertial effects are negligible. Although polymer elasticity delays interface deformation, its contribution to the overall force balance does not exceed approximately 10\%, even for the most elastic fluid investigated. This observation permits further simplification of the model by neglecting inertial and elastic contributions in the droplet-scale momentum balance. The droplet length therefore reduces to $l_d = h/ (C Ca_1)$ or equivalently,

\begin{equation}
l_d = h\left[ \frac{\bar{b}Ca_1}{2(1-\bar{b})^2}
+\frac{\bar{h}Ca_1}{(1-\bar{b})^2}
-\frac{2\bar{h}\beta Ca_2}{\bar{b}(1-\bar{b})}\right]^{-1}.
\label{eqn:12}
\end{equation}

Using the fitted relations for the droplet width obtained from Figure~\ref{fig:3}(d), $\bar{b} = 0.3 Ca_1^{-0.225}$ and $(1-\bar{b}) = 1.6 Ca_1^{0.47}$, the final predictive expression becomes

\begin{equation}
l_d = h \left[ 0.53 Ca_1^{0.47}
+3.55 \bar{h} Ca_1^{0.695}
-4.16 \bar{h} \beta Ca_1^{-0.245} Ca_2 \right]^{-1}.
\label{eqn:13}
\end{equation}

As shown in Figure~\ref{fig:3}(e), this simplified model predicts the experimentally measured primary droplet diameter with nearly the same accuracy as the complete force balance (i.e., equation \eqref{eqn:10}). The model further demonstrates that increasing the continuous-phase capillary number enhances viscous forcing and accelerates pinch-off, resulting in smaller droplets, as demonstrated in Supplementary Movie S2. In contrast, increasing the dispersed-phase capillary number suppresses capillary-driven necking and promotes the formation of larger droplets prior to breakup. These results indicate that, despite the pronounced filament formation characteristic of viscoelastic fluids, the primary droplet size remains governed primarily by the competition between viscous and capillary stresses, with elasticity playing only a secondary role during the initial pinch-off process.

\subsubsection{Cross-stream migration of the primary droplet and filament geometry}
\label{sec:3-2-2}

Following primary pinch-off, the detached droplet does not simply convect downstream but undergoes a pronounced cross-stream migration toward the channel centre while remaining connected to the upstream fluid through a slender viscoelastic filament. This migration determines the position and inclination of the filament immediately before elastocapillary thinning begins and therefore plays a critical role in the subsequent filament evolution discussed in the following sections.

\begin{figure}
\centerline{\includegraphics[width=0.90\columnwidth]{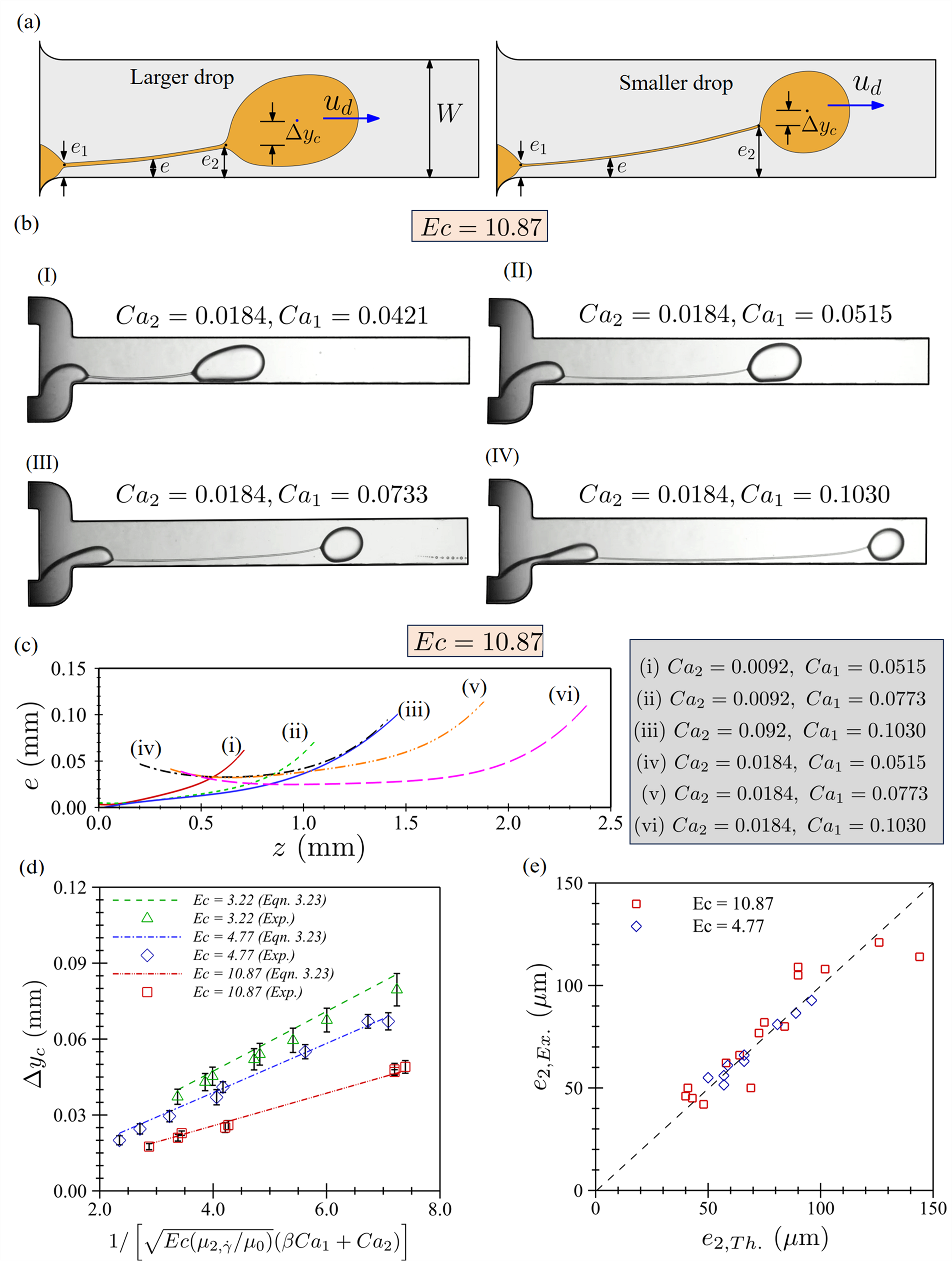}}
   \medskip
  \caption{\justifying{Migration of primary drop and orientation of the filament in the squeezing and dripping regime. (a) Schematic showing the orientation of the filament and position of the migrated droplet. 
(b) (I–IV) Increasing $Ca_1$ (at fixed $Ca_2$) leads to filament elongation and reorientation, which governs its eccentricity ($e$) with respect to the wall for $Ec = 10.87$.
(c) Locus of the filament in the 2D space $(y,z)$: at lower $Ca_2~(=0.0092)$, the tail remains near the wall $(e\approx 0)$ while the head lies near the channel centre, with $e$ increasing along the length [(i)–(iii)]. At higher $Ca_2 (=0.0184)$, the filament orients like a sagging-beam, with the closest point to the wall in between the head and tail [(iv)–(vi)]. (d) Comparison between the experimentally measured $\Delta y_c$ and the scaling prediction (equation~\eqref{eqn:20c})
(e) Comparison of experimentally measured distance of the filament head from the wall $e_{2,Ex.}$ with the theoretical predictions $e_{2,Th.}$ obtained from migration analysis.}}
\label{fig:4}
\end{figure}

Immediately after detachment, the droplet possesses a rounded leading interface connected to the upstream thread through a narrow neck. As the neck collapses into a filament, the droplet gradually migrates away from the channel wall while the upstream attachment point remains anchored near the inlet. Consequently, the filament becomes inclined with respect to the mean flow direction rather than remaining parallel to the channel wall. The downstream end of the filament, indicated by $e_2$, therefore lies farther from the wall than the upstream end marked by $e_1$, producing an asymmetric filament configuration  (see Figure~\ref{fig:4}(a)). The radial offset ($\Delta y_c$) between the droplet centroid and the filament attachment point provides a convenient measure of this migration.

Systematic experiments performed over a wide range of flow conditions reveal two important trends (see Figures~\ref{fig:4}(b) and (c)). First, increasing the continuous-phase Capillary number reduces the primary droplet size while simultaneously increasing the filament length, although the overall inclination changes only weakly. Second, increasing the dispersed-phase Capillary number produces substantially larger lateral displacements, resulting in filament shapes that evolve from gently curved threads to sagging filament configurations whose minimum wall clearance occurs away from either end. These observations indicate that filament geometry is governed primarily by droplet migration rather than by filament stretching alone, as clearly demonstrated by the filament evolution in Supplementary Movie S3.

The observed migration originates from lift forces generated by droplet deformation in confined shear flow. Even in Newtonian systems, deformable droplets experience a deformation-induced non-inertial lift force that drives them toward the channel centreline \citep{Abkarian2002,Chan1979,Rallison1984}. Scaling this lift force as

\begin{equation}
F_{NIL} \sim f_1(1-v) \left(\frac{\mu_1\dot{\gamma}r_p^3}{y}\right), 
\label{eqn:14}
\end{equation}
and balancing it against a Stokes-type lateral drag, $F_L \sim \mu_1 r_p V_m$,
gives the characteristic migration velocity

\begin{equation}
V_{m,NL} \approx Ca^2 \left(\frac{r_p}{W}\right)^2u_{\text{avg}},
\label{eqn:15}
\end{equation}
where $f_1(1-v)$ is a function that characterises the deformability of the droplet, and $v$ denotes the reduced volume, defined as the ratio of the enclosed volume to the droplet surface area. Here, $\mu_1$ is the viscosity of P1, $r_p$ is the effective radius of the droplet, and $y$ is the radial distance from the channel wall to the droplet centre. In \eqref{eqn:15}, $Ca = \mu_1 u_{\text{avg}}/\gamma$, and $u_{\text{avg}} = (Q_1+Q_2)/Wh$ is the average superficial velocity of the two-phase flow inside the channel.   

Because the dispersed phase is viscoelastic, an additional elastic lift force acts on the droplet. Following the analysis of \citet{Hazra2019}, this contribution scales as

\begin{equation}
F_{VE} \sim f_2(1/\beta, Wi_D)\left(\frac{\mu_1 \dot{\gamma} r_p^3}{y}\right),
\label{eqn:16}
\end{equation}
According to \citet{Hazra2019}, when $1/\beta < 3.74$, the viscoelastic lift force satisfies $F_{VE}>0$, causing the droplet to migrate towards the channel centre. In this regime, the function $f_2(1/\beta, Wi_D)$ is well approximated by $f_2 \approx (1/\beta)^{0.33}Wi_D$, from which the corresponding migration velocity is obtained as

\begin{equation}
V_{m,VE} \approx (1/\beta)^{0.33}~Wi_D \left(\frac{r_p}{W}\right)^2 u_{avg},
\label{eqn:17}
\end{equation}
where $\beta~(=3.45)$ is the continuous-to-discrete phase viscosity ratio, and $Wi_D = \lambda_r \dot{\gamma} ~(= 2\lambda_r u_{avg}/W)$ is the Weissenberg number of the discrete phase. 

Combining the Newtonian and viscoelastic contributions gives the characteristic migration velocity

\begin{equation}
V_{m,0} \approx \left[ Ca^2 + (1/\beta)^{0.33}~Wi \right]\left(\frac{r_p}{W}\right)^2 u_{avg}
\label{eqn:18}
\end{equation}
demonstrating that deformation and elasticity contribute additively to the initial lateral motion. The experiments further show that migration progressively weakens as the droplet approaches the channel centreline. To account for this behaviour, the migration velocity is represented by

\begin{equation}
V_m \approx V_{m,0}\left[\left(\frac{y}{W}\right)-A\left(\frac{y}{W}\right)^2\right]
\label{eqn:19}
\end{equation}
where the empirical coefficient obtained from droplet tracking is $A \approx 2.5$. The droplet trajectory is then determined from

\begin{equation}
\frac{dy}{dt} = V_m,
\label{eqn:20}
\end{equation} 
which is integrated over the characteristic migration time. High-speed imaging indicates that the transformation from a connected neck to a fully developed viscoelastic filament occurs over approximately three relaxation times, while most of the lateral migration is completed during the first half of this process. Accordingly, the integration is performed over a characteristic time of approximately $1.5\lambda_r$. The initial droplet position ($y_0$) is estimated from the neck geometry immediately after pinch-off, providing the final radial position of the droplet centroid, $y_n$, prior to the onset of filament instability. A detailed analysis of the cross-stream migration of the primary droplet is presented in \S~S.1 of the Supplementary Material.

Having established the droplet trajectory, the remaining geometric quantity required is the radial offset between the droplet centre and the filament attachment point. Experimental observations show that this offset decreases systematically as the filament becomes longer (see Figure~\ref{fig:4}(b)). Since the filament length increases with both elasticity and imposed shear, whereas the droplet diameter decreases, the attachment point progressively shifts toward the droplet centre, causing the droplet to become more nearly spherical.

Using the scaling for the maximum stable filament length derived later in §3.2.4,

\begin{equation}
l_f \approx 1.5 \frac{h_m W}{\bar{e}} \frac{\mu_{2 \dot{\gamma}}}{\mu_0} Ec (\beta Ca_1+Ca_2)^2
\label{eqn:20a}
\end{equation}
and assuming that the centroid offset varies inversely with both filament length and filament eccentricity, the offset is obtained as

\begin{equation}
\Delta y_c = C_c \left[ 1.5\,h_m\,W~\frac{\mu_{2,\dot{\gamma}}}{\mu_0}\,Ec (\beta \,Ca_1+Ca_2)^2\right]^{-\frac{1}{2}}
\label{eqn:20c}
\end{equation}
where $C_c$ is an empirical constant determined from the experiments, $h_m$ denotes the hydraulic diameter of the channel, and $\bar{e}$ represents the mean clearance between the filament and the channel wall.

The filament position is then obtained directly from the predicted droplet trajectory, $e_2 = y_n - \Delta y_c$. Figures~\ref{fig:4} (d) and (e) demonstrate that the proposed model accurately predicts both the centroid offset and the final filament eccentricity over the complete range of Capillary numbers and elasticities investigated. The data collapse obtained using the expression for $\Delta y_c$ indicates that filament positioning is governed primarily by the coupled effects of droplet migration and viscoelastic stretching rather than by geometric confinement. 
This result has important consequences for the subsequent filament dynamics. The eccentricity predicted here determines the local wall-induced shear acting on the filament and therefore directly enters the scaling analyses developed in \S\ref{sec:3-2-3} and \S\ref{sec:3-2-4} for predicting the critical filament diameter and the maximum stable filament length. Consequently, primary droplet migration is not merely a geometric consequence of breakup but establishes the initial conditions governing the entire filament-thinning process.

\subsubsection{Scaling of the critical filament thickness}
\label{sec:3-2-3}

Following primary droplet detachment, the connecting viscoelastic filament continues to elongate under the action of the surrounding flow while its radius progressively decreases. The filament-thinning dynamics up to the onset of capillary breakup are examined in Appendix~\ref{appE}. Breakup does not occur immediately after filament formation but only after the filament reaches a critical thickness, beyond which capillary instability overcomes the stabilising effect of polymer elasticity. Predicting this critical thickness is therefore essential for understanding the onset of filament fragmentation and the subsequent formation of secondary droplets.

Unlike unconfined extensional flows \citep{EggersHerrada2020}, the filament in the present confined coflow is continuously subjected to shear imposed by the surrounding continuous phase. Consequently, the filament dynamics are governed not only by the classical balance between capillary and elastic stresses, but also by wall-mediated interfacial shear. This additional mechanism constitutes one of the principal differences between the present confined configuration and freely suspended viscoelastic filaments.

\begin{figure} \centerline{\includegraphics[width=0.98\columnwidth]{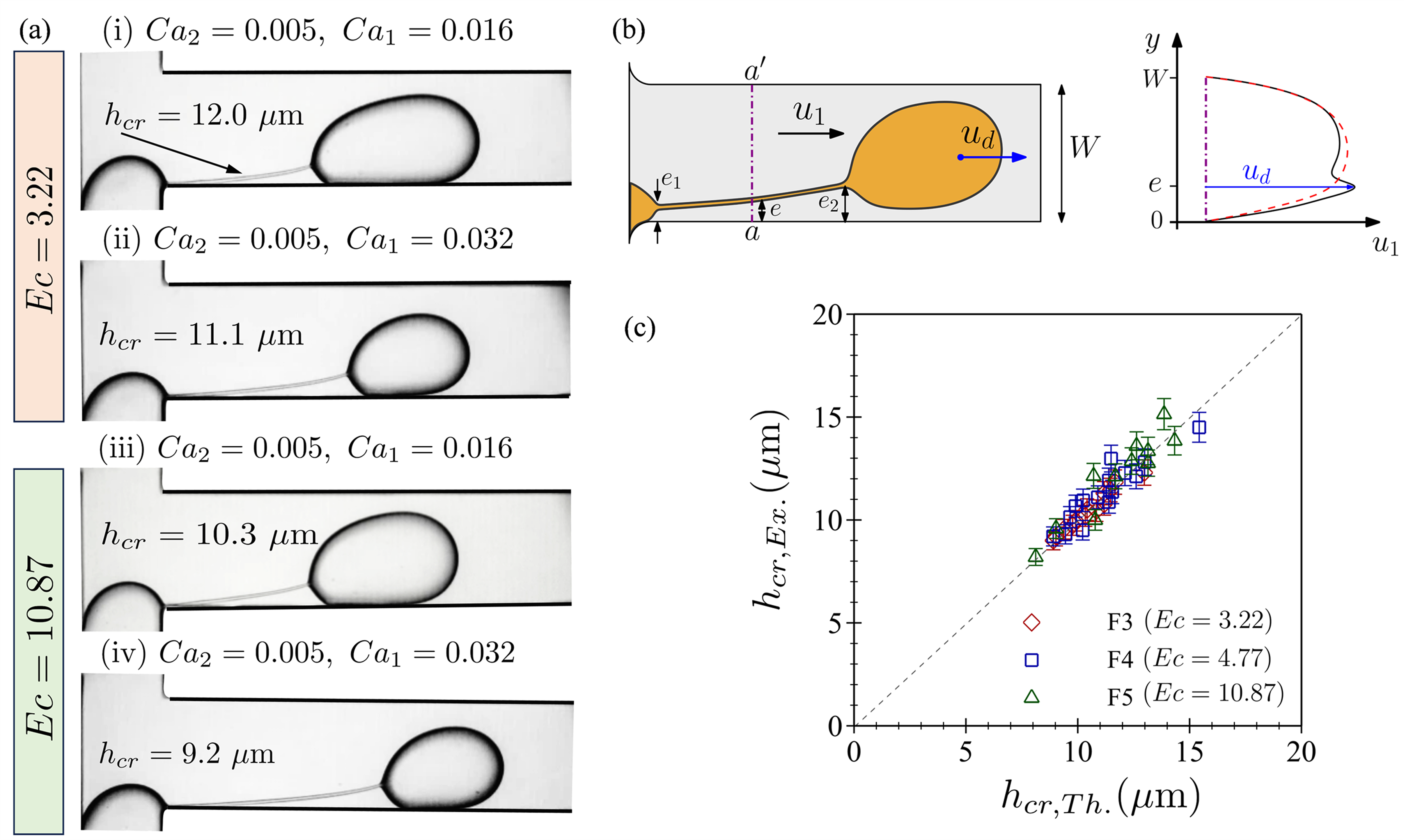}}
   \medskip
  \caption{\justifying{Critical filament thickness prior to instability. (a) Experimental images illustrating the dependence of the critical filament thickness $h_{cr}$ on $Ca_1$, $Ca_2$, and $Ec$. (b) Schematic showing the filament orientation immediately before breakup and the corresponding modification of the hypothetical axial velocity profile of the continuous phase induced by filament motion. (c) Comparison of the theoretically predicted critical filament thickness, $h_{cr,Th.}$, obtained from \eqref{eqn:21}, with the corresponding experimentally measured values, $h_{cr,Ex.}$. }}
\label{fig:5}
\end{figure}

Several important physical trends emerge from the experiments (see Figure~\ref{fig:5}(a)). At fixed $Ca_2$, increasing the continuous-phase Capillary number reduces the critical filament thickness. Stronger continuous-phase shear increases the interfacial strain rate, thereby enhancing polymer stretching and delaying capillary instability until substantially thinner filaments are reached. Likewise, for fixed flow conditions, increasing the elastocapillary number strengthens the elastic tensile stress, requiring larger capillary pressures before instability can develop and therefore reducing $h_{cr}$. Consequently, both increasing flow strength and increasing elasticity promote extended filament thinning prior to breakup.

Immediately prior to breakup, the filament undergoes quasi-steady thinning, during which the capillary pressure generated by the curved interface is balanced by the tensile stress arising from polymer stretching. Following the Oldroyd-B approximation adopted in the previous section, the elastic tensile stress is estimated as $\sigma_e \approx 2\mu_p \lambda_r \dot{\gamma}^2$, where $\mu_p=\mu_0-\mu_s$ denotes the polymeric contribution to the solution viscosity and $\dot{\gamma}$ is the interfacial strain rate. Balancing this stress against the capillary pressure inside the filament yields

\begin{equation}
2 \mu_p \lambda_r \dot{\gamma}^2 \approx \frac{2\gamma}{h_{cr}}
\label{eqn:21}
\end{equation}
where $h_{cr}$ is the critical filament thickness immediately preceding the onset of instability.

The principal challenge is therefore the estimation of the interfacial strain rate. Figure~\ref{fig:5}(b) illustrates the physical mechanism responsible for filament stretching. After pinch-off, the primary droplet is convected downstream with an axial velocity comparable to the superficial mixture velocity, $ u_d = u_{avg} = (Q_1+Q_2)/(Wh)$, which exceeds the mean velocity of the continuous phase (see Figure~\ref{fig:5}(b)). Owing to the no-slip condition at the interface, the surrounding continuous phase is locally accelerated near the filament, producing a strong velocity gradient between the wall and the moving interface. Unlike conventional extensional thinning, the deformation is therefore driven primarily by wall-induced shear.

Assuming that the velocity varies approximately linearly across the gap separating the filament from the channel wall, the characteristic strain rate scales as $\dot{\gamma} \sim u_d/e$, where $e$ denotes the local filament eccentricity. Since the filament is inclined with respect to the wall, the wall clearance varies continuously along its length. Using the migration model developed in \S\ref{sec:3-2-2}, the average wall clearance is approximated as $\bar{e} \sim e_2/4$, which gives the effective interfacial strain rate $\dot{\gamma} \approx 4u_d/e_2$. The shear-dependent polymer viscosity is then evaluated from the Carreau–Yasuda model (equation~\eqref{eqn:3.1}), and substituted into \eqref{eqn:21} to determine the critical filament thickness. The resulting predictions are compared with experiments in Figure~\ref{fig:5}(b). For each fluid pair, the critical thickness was measured immediately before the appearance of the first capillary undulations along the filament, with each data point representing the average of three independent measurements. The model captures the experimental trends over the complete range of operating conditions.

Figure~\ref{fig:5}(c) compares the experimentally measured and predicted critical filament thicknesses for all fluid pairs investigated. The data collapse closely around the line of unity slope, demonstrating that the proposed stress balance successfully captures the onset of filament instability despite the simplifying assumptions used in estimating the interfacial strain rate.

The present analysis also highlights an important physical distinction from previous theories of viscoelastic filament thinning. Existing models for freely suspended filaments generally assume that capillary pressure is balanced solely by elastic tension immediately prior to breakup \citep{EggersHerrada2020}. In confined microchannels, however, polymer stretching is generated by shear transmitted through the surrounding continuous phase. The resulting wall-mediated interfacial shear provides the missing physical link between the imposed flow conditions and filament stability, thereby explaining why the critical filament thickness depends explicitly on the Capillary numbers in addition to the elastocapillary number. Furthermore, analysis of the filament-thinning dynamics prior to the onset of instability provides an independent estimate of the polymer relaxation time. As demonstrated in Appendix~\ref{appE}, the estimated relaxation time is only weakly dependent on the capillary-number ratio ($Ca_r$), indicating that the proposed approach is robust over the range of flow conditions investigated.

\subsubsection{Scaling of the maximum stable filament length}
\label{sec:3-2-4}

Following primary droplet formation, the viscoelastic thread connecting the droplet to the upstream liquid is continuously stretched by the downstream motion of the primary droplet, resulting in progressive filament thinning until the onset of instability and eventual breakup. Unlike Newtonian fluids, where capillary forces rapidly drive pinch-off, viscoelastic filaments remain stable over considerably longer distances because polymer stretching continuously generates tensile stresses that oppose capillary contraction. Consequently, the maximum filament length immediately prior to breakup provides a direct measure of the competition between viscous stretching, capillary relaxation, and polymer elasticity.

\begin{figure}  \centerline{\includegraphics[width=0.95\columnwidth]{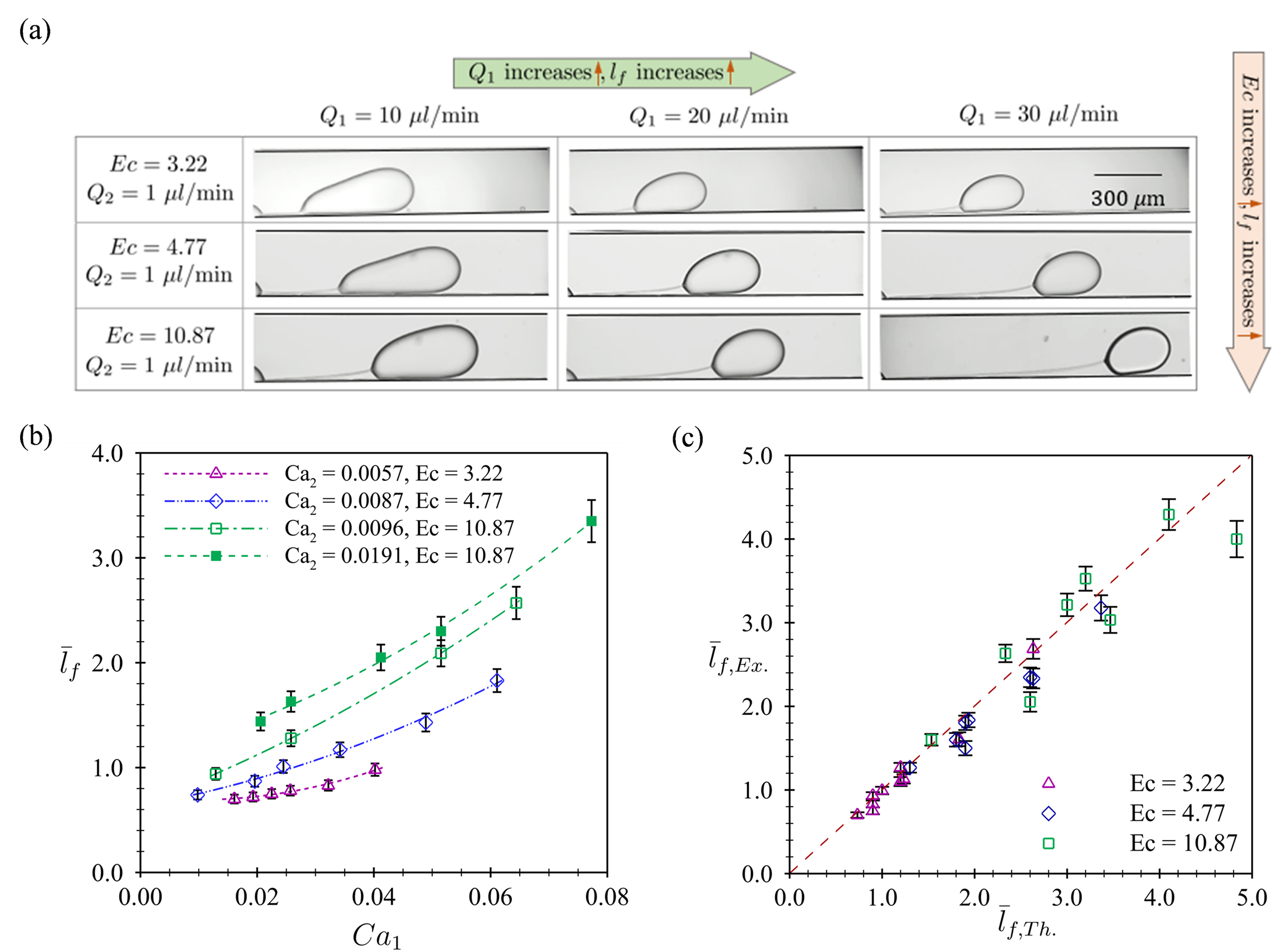}}
   \medskip
  \caption{\justifying{Maximum filament length $l_f$ at the onset of breakup.
(a) Experimental images showing that $l_f$ increases with increasing continuous and dispersed phase flow rates ($Q_1$, $Q_2$) and relaxation time $\lambda_r$ (or $Ec$).
(b) Variation of the dimensionless filament length $\bar{l}_f = l_f/W$ with $Ca_1$ for fixed $Ca_2$ and $Ec$.
(c) Comparison of experimentally measured filament lengths $\bar{l}_{f,Ex.}$ with theoretical predictions $\bar{l}_{f,Th.}$ from \eqref{eqn:23}.}}
\label{fig:6}
\end{figure}

Representative filament evolution for different flow conditions is shown in Figure~\ref{fig:6}(a). At fixed dispersed-phase flow rate and relaxation time, increasing the continuous-phase flow rate produces progressively longer filaments owing to stronger viscous stretching. Likewise, for fixed flow rates, increasing the polymer relaxation time substantially increases the filament length, indicating that elasticity delays capillary breakup by sustaining tensile stresses within the filament. These observations demonstrate that filament extension is controlled jointly by the imposed shear and the elastic response of the dispersed phase.

The corresponding quantitative measurements are summarized in Figure~\ref{fig:6}(b). For fixed $Ca_2$, the dimensionless filament length increases approximately linearly with $Ca_1$, while the slope increases systematically with the elastocapillary number $E_c$. Conversely, at fixed elasticity, increasing $Ca_2$ produces an approximately uniform upward shift of the data without significantly altering the slope. These trends suggest that the continuous-phase Capillary number primarily governs the stretching rate, whereas the dispersed-phase Capillary number determines the initial level of deformation. The enhancement associated with increasing elasticity reflects the longer persistence of polymer stresses during filament stretching.

To develop a predictive scaling, we consider the characteristic lifetime of a thinning viscoelastic filament. For Newtonian fluids, breakup occurs over the viscocapillary time scale $t_c = h_m\mu_2/\gamma$, where $h_m$ is the hydraulic diameter of the channel. In viscoelastic fluids, however, filament thinning is delayed because the polymer relaxation time becomes comparable to the deformation time. Motivated by this competition, we approximate the characteristic thinning time as $t_{ve} = t_c (1+Wi_c) = 1.5 h_m (\mu_{2, \dot{\gamma}} / \gamma) (1+Wi_c)$, where $Wi_c = \lambda_r \dot{\gamma}$ is the Weissenberg number evaluated immediately before breakup. This expression naturally recovers the Newtonian limit ($Wi_c \rightarrow 0$) while accounting for the additional lifetime associated with elastic stress development. The interfacial strain rate $\dot{\gamma}$ is evaluated using the wall-induced shear model developed in \S\ref{sec:3-2-3}.

Since the primary droplet translates downstream at an approximately constant velocity, $ u_d = (Q_1+Q_2)/(Wh)$, the maximum filament length follows directly as $l_f = t_{ve}u_d$, which gives
\begin{equation}
l_f = 1.5 \, \frac{h_m \mu_{2\dot{\gamma}}}{\gamma}\, (1+Wi_c) \, u_d.
\label{eqn:22}
\end{equation} 
Expressing the Weissenberg number in terms of the wall-induced strain rate and using the definitions of the elastocapillary and Capillary numbers yields

\begin{equation}
\frac{l_f}{W} = \bar{l}_f= 1.5 \,\frac{h_m}{\bar{e}} \frac{\mu_{2,\dot{\gamma}}}{\mu_0}\,Ec (\beta \,Ca_1+Ca_2)^2
\label{eqn:23}
\end{equation}

which predicts that the maximum filament length is governed by the coupled effects of confinement, elasticity, viscosity ratio, and the imposed shear.

Equation~\eqref{eqn:23} further predicts $d\bar{l}_f/dCa_1 = 2\,A \,\beta\,Ec\, (\beta Ca_1 + Ca_2)$, where $A = (h_m/\bar{e})(\mu_{2,\dot{\gamma}}/\mu_0)$. The slope of the filament-length curves therefore increases linearly with the elastocapillary number, consistent with the experimental observations in Figure~\ref{fig:6}(b). Physically, increasing elasticity amplifies the tensile stresses generated by a given imposed shear, thereby allowing the filament to remain stable over progressively longer distances before capillary instability develops. The theoretical predictions are compared with the experimental measurements in Figure~\ref{fig:6}(c). The data collapse closely around the line of equality, demonstrating that the proposed scaling captures the maximum filament length over the complete range of Capillary numbers and elasticities investigated.

The present scaling also provides a physical interpretation of the observed filament extension. Increasing $Ca_1$ enhances the wall-induced shear acting on the filament, increasing the rate of polymer stretching. Increasing $Ca_2$ supplies a larger volume of viscoelastic fluid to the growing thread, producing longer filaments before breakup. Finally, increasing the elastocapillary number prolongs the lifetime of elastic stresses, allowing the filament to resist capillary contraction for a longer period. The maximum filament length is therefore determined not by elasticity alone, but by the coupled action of viscous stretching, capillary relaxation, and polymer stress development under confined shear. This interpretation is further supported by the filament-thinning dynamics presented in Appendix~\ref{appE}, which show that, even for a fixed elastic property, the thinning behaviour depends on the capillary-number ratio ($Ca_r$), reflecting the influence of the relative viscous stresses exerted by the two phases.

\subsection{Critical thread length and primary droplet size in the jetting regime }
\label{sec:3-3}

For sufficiently elastic fluids, the breakup dynamics undergo a qualitative transition from dripping to jetting. Unlike the squeezing and dripping regimes, where breakup occurs immediately downstream of the junction, the dispersed phase forms a continuous viscoelastic thread that extends over a finite distance before becoming unstable. Within the present parameter range, jetting is observed only for the moderately and highly elastic fluids ($E_c=3.22, 4.77$ and $10.87$), whereas the weakest elastic fluid ($E_c=1.28$) exhibits only squeezing and dripping. The onset of jetting occurs when the dispersed-phase capillary number exceeds approximately $Ca_{2,\dot{\gamma}} \approx 0.018$.

\begin{figure}  \centerline{\includegraphics[width=0.90\columnwidth]{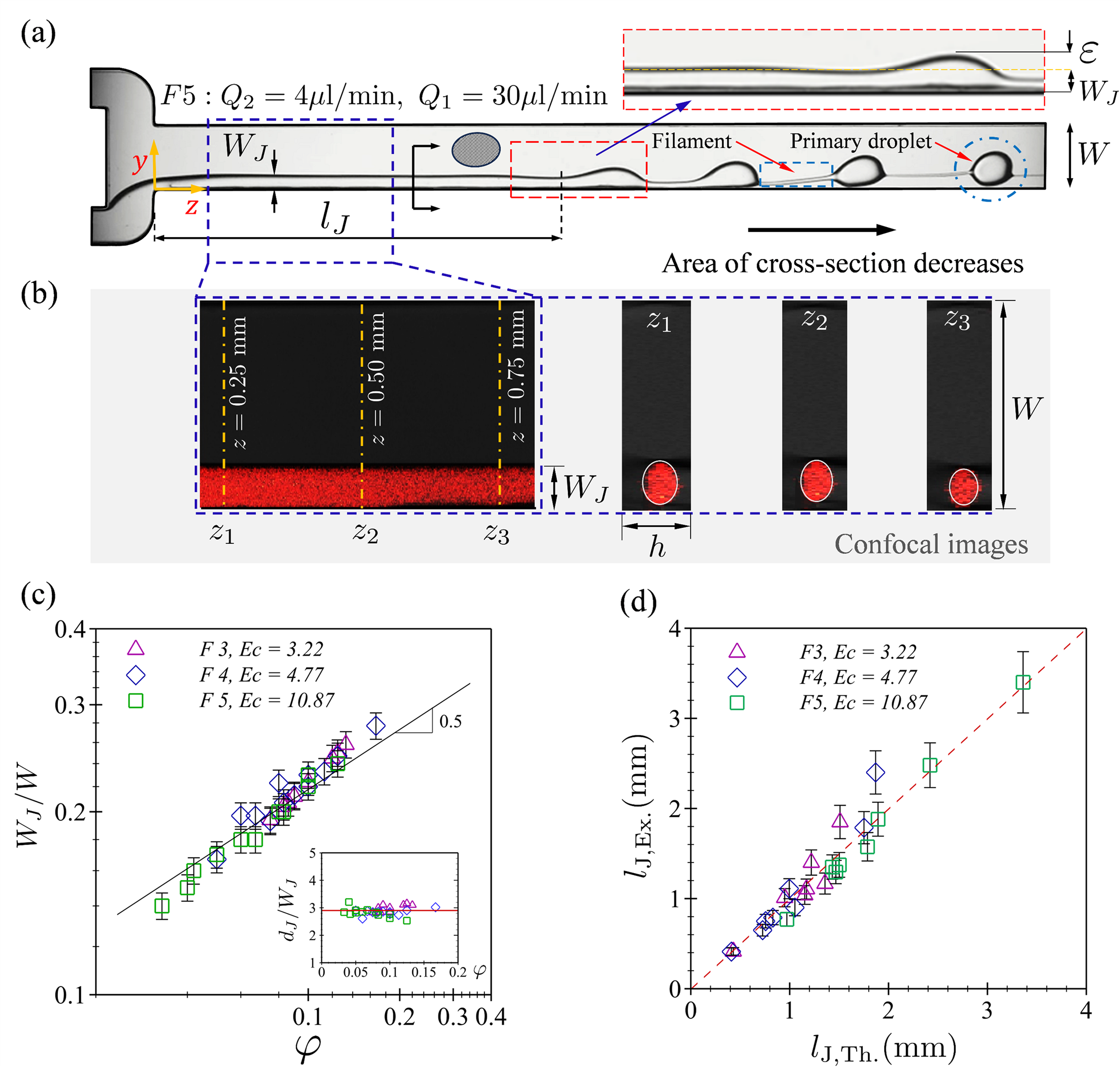}}
   \medskip
  \caption{\justifying{Critical thread length $l_J$, jet width $W_J$, and primary droplet diameter $d_J$ in the jetting regime. (a) Experimental image of fluid pair $F5$ (Table~\ref{tab:table2}) at $Q_2 = 4~\mu\text{l/min}$ and $Q_1 = 30~\mu\text{l/min}$, illustrating the jetting regime and indicating the jet width, $W_J$, and thread length, $l_J$. (b) Confocal microscopy images showing the progressive reduction in the thread cross-sectional area along the downstream direction. (c) Variation of the normalised jet width, $W_J/W$, with the flow-rate ratio, $\varphi$. The solid line represents the fitted correlation given by Eqn.~\eqref{eqn:24}. The inset shows that the ratio $d_J/W_J$ is independent of $\varphi$; the solid line corresponds to $d_J/W_J = 2.9$. (d) Comparison of the experimentally measured thread lengths, $l_{J,\mathrm{Ex.}}$, with the corresponding theoretical predictions, $l_{J,\mathrm{Th.}}$, obtained from Eqn.~\eqref{eqn:26}.}}
\label{fig:7}
\end{figure}

Figure~\ref{fig:7}(a) illustrates a representative jetting event. A nearly cylindrical thread emerges from the junction and remains stable over an unperturbed length before capillary disturbances amplify and eventually produce a sequence of primary droplets connected by thin viscoelastic filaments. Confocal images acquired at successive downstream locations (see Figure~\ref{fig:7}(b)) further show that the thread cross-section remains only slightly elliptical while gradually decreasing in area, indicating continuous axial stretching of the dispersed phase. Consequently, the axial velocity increases downstream owing to mass conservation, promoting further extensional deformation prior to breakup.

The observed breakup mechanism closely resembles the capillary instability of confined viscous jets reported by \citep{CubudTMason2008}. The principal distinction is that, in the present system, polymer elasticity delays the growth of capillary disturbances. For Newtonian confined coflows, jetting typically requires a large viscosity contrast to suppress premature breakup. In contrast, the present fluids exhibit jetting despite a relatively modest viscosity ratio ($\beta^{-1} = 45$), demonstrating that elastic tensile stresses provide an additional stabilising mechanism that enables the thread to survive over substantially longer distances before breakup.

The width of the unperturbed thread immediately downstream of the junction was measured for all operating conditions. Figure~\ref{fig:7}(c) shows that the thread width depends only on the flow-rate ratio and is essentially independent of elasticity, yielding
\begin{equation}
\frac{W_J}{W} \approx \sqrt{\frac{Q_2}{2Q_1}}
\label{eqn:24}
\end{equation}
This scaling is identical to that reported by \citep{CubudTMason2008} for confined Newtonian coflows, suggesting that elasticity has little influence on the initial focusing of the dispersed stream.

The diameter of the primary droplet generated from the jet exhibits an analogous dependence,
\begin{equation}
\frac{d_J}{W} \approx 2.9\sqrt{\frac{Q_2}{2Q_1}}
\label{eqn:25}
\end{equation}
indicating that both the thread diameter and the primary droplet size are governed primarily by the hydrodynamic focusing imposed by the flow-rate ratio. Combining the experimentally fitted correlation, \eqref{eqn:24}, with \eqref{eqn:25} yields $d_J/W_J \approx 2.9$. This value is in excellent agreement with the direct experimental measurements shown in the inset of figure~\ref{fig:7}$(c)$. This ratio is slightly smaller than the value reported for Newtonian jets ($\approx3.1$) by \citep{CubudTMason2008}. The reduction arises because approximately $5$\% of the dispersed-phase volume remains within the connecting viscoelastic filament and is subsequently converted into satellite droplets rather than contributing to the primary droplet volume.

The measured droplet volume also provides an estimate of the wavelength of the fastest-growing disturbance. Assuming conservation of volume, $1.05 \pi d_J^3/6\approx\lambda ,\pi W_J^2/4$, from which the dimensionless disturbance wavenumber becomes $x = \pi W_J/\lambda=0.184 \pm 0.055$. 

Remarkably, despite varying the polymer relaxation time by more than one order of magnitude, the measured wavenumber remains essentially constant and closely matches the value reported for Newtonian confined jets \citep{CubudTMason2008}. These observations suggest that elasticity has little influence on wavelength selection and instead primarily modifies the temporal amplification of disturbances. This behaviour is consistent with classical linear stability theory. For viscous jets, disturbances evolve according to $\varepsilon(t) = \varepsilon_0 e^{\omega t}$, with the capillary growth rate scaling as $\omega \sim (\gamma / \mu_2 R_J)(1 - k^2 R_J^2)$. Because polymer stretching resists extensional deformation, we propose that elasticity reduces the growth rate according to $\omega \sim (\gamma / \mu_2 R_J)(1 - k^2 R_J^2)/(1+Wi)$. As the instability propagates downstream, its amplitude grows with time, eventually leading to the formation of the primary droplet. The disturbance amplitude, $\varepsilon$, increases rapidly as the jet approaches the breakup point, ultimately causing jet breakup (see Figure~\ref{fig:7}$a$). For example, at a fixed axial location, let the disturbance amplitude increase from $\varepsilon_{1}$ to $\varepsilon_{2}$ over a time interval $\Delta t$. Assuming that the disturbance grows exponentially with a constant growth rate, $\omega$, the amplitude satisfies $\varepsilon_{2}/\varepsilon_{1}=\exp(\omega \Delta t)$. Accordingly, the disturbance growth rate can be determined directly from the experimental images as $\omega=(1/\Delta t)\ln(\varepsilon_{2}/\varepsilon_{1})$. The experimentally measured growth rates closely follow the proposed scaling, indicating that increasing fluid elasticity suppresses disturbance amplification while leaving the dominant instability wavelength nearly unchanged.

The reduced growth rate directly determines the distance over which the thread remains stable. For Newtonian confined jets, breakup occurs over the viscocapillary time scale $t_c \sim W_J \mu_2/\gamma$ \citep{CubudTMason2008,Eggers2008}. The present experiments demonstrate that elasticity increases this characteristic time approximately in proportion to $Wi/Ca_{2,\dot{\gamma}}$, giving $T_c = t_c Wi/Ca_{2,\dot{\gamma}}$. Since disturbances are convected downstream with the mean jet velocity, $u_2 = 4Q_2/\pi W_J^2$, the critical thread length becomes

\begin{equation}
l_{J,Th.}\approx 4 ~\frac{\mu_{2,\dot{\gamma}}}{\gamma}\frac{Wi}{\pi W Ca_{2,\dot{\gamma}}}\sqrt{\frac{Q_1Q_2}{2}}.
\label{eqn:26}
\end{equation}

Figure \ref{fig:7}(d) compares the predicted thread length with measurements for all three elastic fluids. The agreement is consistently good over the entire experimental range, indicating that the proposed scaling captures the dominant balance between capillary instability, viscous dissipation and polymer elasticity governing the onset of jet breakup.

\subsection{Filament dynamics in squeezing and dripping regimes}
\label{sec:3-4}

\begin{figure}
  \centerline{\includegraphics[width=0.95\columnwidth]{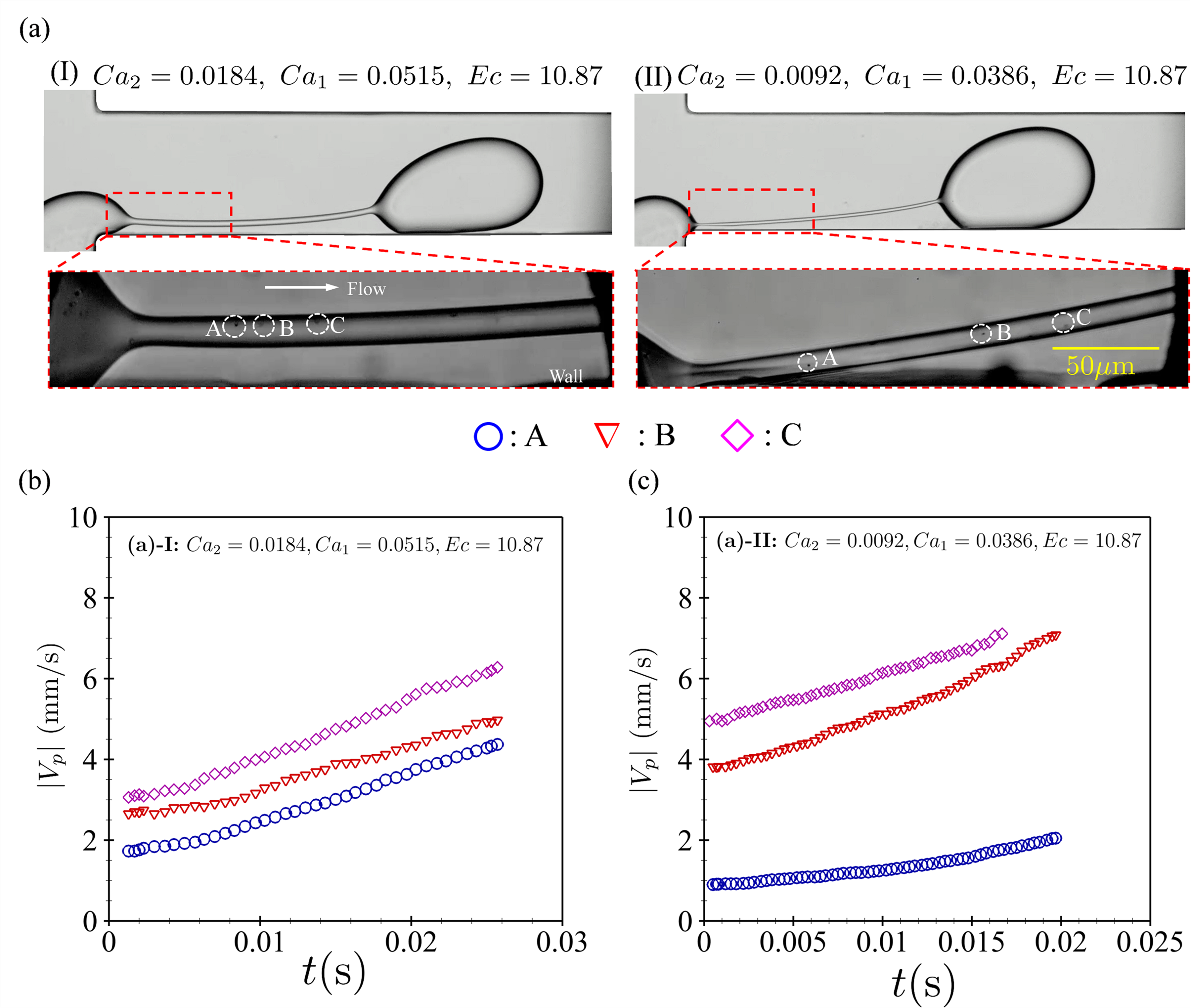}}
   \medskip
  \caption{\justifying{Effect of filament inclination on the internal fluid velocity for a highly elastic fluid ($Ec = 10.87$), quantified by tracking $1\,\mu\mathrm{m}$ tracer particles. 
(a) Experimental images of a stable filament: (I) at $Ca_2 = 0.0184$ and $Ca_1 = 0.0515$, where a portion of the filament remains parallel to the wall near the junction; the magnified view ($60\times$) indicates the initial positions of the tracked particles. (II) At $Ca_2 = 0.0092$ and $Ca_1 = 0.0386$, the filament becomes inclined; the zoomed view shows three particles located at different heights from the wall. 
(b) Comparison of the velocity magnitude of particles (\textbf{A},\textbf{B} and \textbf{C}) moving at nearly the same distance from the wall, showing only small deviations from the mean value (velocity of particle \textbf{B}, $|V_P|_B$). 
(c) Velocity of a particles (\textbf{B,C}) travelling within a region of the filament farther from the wall, exhibiting a higher velocity (4 and 5 times) compared to those closer to the wall (\textbf{A}).}}
\label{fig:8}
\end{figure}

Following the formation of the primary droplet, a thin viscoelastic filament (diameter, $h_f \sim 10~\mu$m) connects the droplet to the upstream liquid until capillary instability eventually causes breakup. As demonstrated in \S\ref{sec:3-2-2}, the filament does not always remain parallel to the channel wall; depending on the flow conditions, it may either remain nearly parallel over part of its length or become fully inclined (Figure~\ref{fig:8}(a)). Since the distance between the filament and the wall varies along an inclined filament, the surrounding continuous phase imposes a spatially non-uniform shear field. This raises an important question: does the interfacial velocity remain uniform along the filament, or does wall-induced shear generate local variations in the internal flow?

To address this question, we performed particle-tracking experiments by dispersing $1.0~\mu$m polystyrene microspheres within the viscoelastic phase. As shown in \S\ref{sec:3-2-4}, highly elastic fluids produce substantially longer filaments, thereby providing sufficient observation time for tracking individual particles. Consequently, the experiments presented here were conducted using fluid $F5$ ($Ec = 10.87$), as listed in Table~\ref{tab:table1}.

Figure~\ref{fig:8}(a) illustrates two representative filament configurations. Case I corresponds to a partially inclined filament obtained at $Ca_1=0.0515$ and $Ca_2=0.0184$, whereas Case II represents a fully inclined filament generated at $Ca_1=0.0386$ and $Ca_2=0.0092$. Three tracer particles (A, B and C) are identified within each filament and their trajectories are monitored throughout filament stretching. In Case I, the three particles are located at nearly identical distances from the channel wall, whereas in Case II, they occupy distinctly different wall-normal positions. The measured particle velocities are presented in Figures~\ref{fig:8}(b,c). For the nearly horizontal filament (Figures~\ref{fig:8}(b)), the velocities of particles A, B and C increase continuously as the filament thins. This acceleration is expected because capillary thinning increases the axial pressure gradient within the filament, thereby accelerating the internal flow. Since all three particles experience nearly identical local confinement, their velocities remain comparable throughout the stretching process. The small differences observed between individual trajectories are attributed primarily to slight variations in particle depth relative to the imaging plane, as indicated by the weak defocusing of particles B and C. A markedly different behaviour is observed for the inclined filament (Figure~\ref{fig:8}(c)). Here, particle A, located closest to the channel wall, consistently moves more slowly than particles B and C, with its velocity reduced by approximately a factor of four to five. The reduction arises because the neighbouring continuous phase experiences stronger viscous resistance close to the wall, producing a larger interfacial shear stress that retards the local filament motion. Particle C exhibits a slightly higher velocity than particle B because it is both farther from the wall and marginally farther from the interface, thereby experiencing weaker viscous resistance. Consequently, the particle velocity increases systematically with increasing distance from the wall.

These measurements provide direct experimental evidence that an inclined filament does not translate as a rigid body. Instead, the varying wall clearance produces a spatially varying interfacial shear stress, leading to significant variations in the local axial velocity within the filament, even though the filament itself is advected downstream with an approximately constant mean velocity $u_d$. Similar behaviour is observed for two additional combinations of $Ca_1$ and $Ca_2$, confirming that the observed velocity gradients are a robust consequence of wall-induced confinement rather than a particular operating condition. The corresponding measurements are presented in \S~S.2 of the Supplementary Material.


A defining characteristic of viscoelastic liquids is their ability to sustain long-lived, highly stretched filaments, in contrast to Newtonian filaments that undergo relatively rapid capillary breakup \citep{Anna2001}. Nevertheless, these viscoelastic filaments eventually lose stability under continued stretching and develop the characteristic beads-on-a-string (BOAS) morphology reported in previous studies \citep{WagnerEggers2005,SalterEggers2008,ArdekaniSharma2010}. An important question, however, is where the instability first develops in a confined microchannel and whether wall-induced confinement influences the initiation of breakup.

\begin{figure}
  \centerline{\includegraphics[width=1.0\columnwidth]{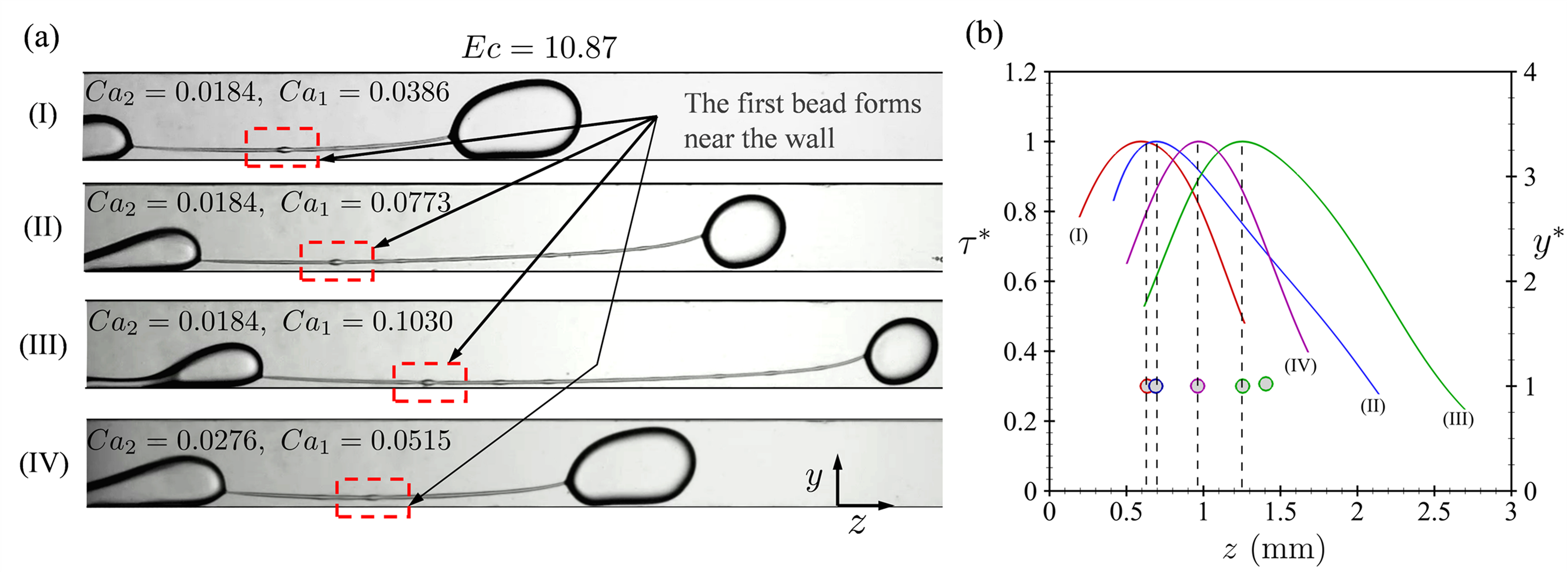}}
   \medskip
  \caption{\justifying{Bead formation during filament thinning at different capillary numbers. 
(a) Panels I--IV present experimental images for various capillary number combinations, showing that the first bead consistently forms in the region of the filament closest to the wall. 
(b) Distribution of the effective shear stress exerted by the surrounding fluid along the filament, demonstrating that the first bead emerges at the location of maximum shear stress.}}
\label{fig:9}
\end{figure}

Figure~\ref{fig:9}(a) addresses this question using a highly elastic fluid ($E_c=10.87$). The experiments reveal a remarkably robust behaviour: irrespective of the operating conditions, the first observable perturbation always develops at the portion of the filament closest to the channel wall. In cases I-III, $Ca_1$ is increased from $0.0386$ to $0.1030$ while maintaining $Ca_2=0.0184$. As expected from the scaling developed in \S\ref{sec:3-2-4}, increasing $Ca_1$ produces a smaller primary droplet and a substantially longer filament before breakup. Despite these significant changes in the global filament geometry, the earliest visible perturbation, identified by the first bead highlighted in Figure~\ref{fig:9}(a), consistently nucleates at the wall-side portion of the filament. The same behaviour is observed for a moderately elastic fluid ($E_c=4.77$), as demonstrated in Supplementary Movie S4. The robustness of this observation is further confirmed in Case IV, where both $Ca_1$ and $Ca_2$ are increased simultaneously. Although the increased capillary numbers modify both the filament length and the overall stretching dynamics, the first bead again appears at the location nearest the wall. These observations demonstrate that the onset of instability is not governed by the global flow rate or the overall filament length, but rather by the local hydrodynamic environment imposed by wall confinement. To examine this mechanism quantitatively, Figure~\ref{fig:9}(b) compares the measured bead locations with the calculated distribution of the normalised interfacial shear stress, $\tau^{\ast} = \tau/\tau_{\text{max}}$, along the filament. The bead locations are represented by their normalised wall-normal positions, $y^{\ast} = y/y_{\text{min}}$, where $y_{\text{min}}$ denotes the minimum wall clearance. For every operating condition investigated, the first bead forms at $y^{\ast} = 1$, corresponding precisely to the point of minimum wall clearance. Furthermore, the vertical reference lines intersect the corresponding shear-stress distributions at $\tau^{\ast} = 1$, demonstrating that bead nucleation always coincides with the location of maximum interfacial shear.

These results provide direct experimental evidence that wall-induced shear determines the location at which instability first develops. The confinement produces a highly non-uniform shear distribution along the inclined filament, generating local stress concentrations that accelerate the amplification of capillary disturbances near the wall. Thus, although capillary forces ultimately drive filament breakup, the initiation of the instability is strongly controlled by the spatial variation of the surrounding shear field. Having established that wall-induced shear governs the onset of instability, the next question concerns the subsequent evolution of the disturbance. Specifically, it remains to be determined whether the repeated bead formation observed in confined viscoelastic filaments can still be interpreted within the classical Rayleigh--Plateau instability (RPI) framework, or whether elasticity fundamentally alters the instability mechanism. This issue is examined in the following section.

The localisation of the first bead at the region of maximum wall-induced shear establishes where the instability is initiated. However, the subsequent evolution of the filament involves multiple generations of capillary instability rather than a single breakup event. After the first perturbation develops, the filament remains connected by thin liquid bridges, which continue to stretch and thin. These connecting threads subsequently undergo secondary and tertiary instabilities, producing progressively smaller beads through repeated sinusoidal modulation of the interface until complete filament rupture is achieved. This hierarchical beads-on-a-string (BOAS) evolution is a characteristic feature of viscoelastic filament breakup. Such behaviour differs fundamentally from Newtonian filament breakup, where a single generation of droplets is generally produced through the classical Rayleigh$-$Plateau instability (RPI) \citep{Eggers2014}. Previous studies have further suggested that even the first-generation instability of viscoelastic filaments departs from the classical RPI because the beads remain connected by long-lived filament bridges \citep{WagnerEggers2005,ChenAshgriz2025}. Nevertheless, it remains unclear whether the evolution of successive generations can still be interpreted within the Rayleigh$-$Plateau framework. This question is examined below.

The extension of the classical Rayleigh analysis to viscous liquid cylinders predicts that the maximum growth rate of long-wave disturbances is \citep{Weber1931,chandrasekhar2013hydro}

\begin{equation}
\omega_{\text{max}} = \left[\left(\frac{\rho_{2}h_{\text{f}}^3}{\gamma}\right)^{1/2}+\frac{3\mu_2 h_{\text{f}}}{\gamma}\right]^{-1},
\label{eqn:27}
\end{equation}

while the corresponding dominant wavelength is

\begin{equation}
\lambda_{\text{max}} = \pi  \left[h_{\text{f}}^2+3\mu_2h_{\text{f}}\sqrt{\frac{h_{\text{f}}}{\rho_2\gamma}}\right]^{1/2}.
\label{eqn:28}
\end{equation}

Recently, \citep{ChenAshgriz2025} examined the stretching and breakup of a viscoelastic liquid bridge and evaluated the applicability of these classical expressions. They showed that using the zero-shear viscosity in  equations  \eqref{eqn:27} and \eqref{eqn:28} predicts the instability wavelength reasonably well but overestimates the growth rate by almost two orders of magnitude. Conversely, replacing the zero-shear viscosity with the extensional viscosity substantially improves the prediction of the growth rate while producing wavelengths that are more than one order of magnitude larger than the experimental values. These observations demonstrate that classical RPI theory cannot be directly extended to viscoelastic filaments without incorporating the underlying rheological response. The present configuration differs fundamentally from that investigated by \citep{ChenAshgriz2025}. Here, the filament evolves within a confined microchannel while simultaneously experiencing capillary forces, wall-induced shear and viscoelastic stresses generated by the surrounding fluid. Consequently, rather than adopting either the zero-shear or extensional viscosity, we formulate an effective viscosity based on the dominant stress balance governing the late stages of filament thinning.

Following \citep{EggersHerrada2020}, the final stage of thinning is dominated by the axial polymeric normal stress, whereas the remaining stress components become negligible (i.e., $\sigma_{rr}$, $\sigma_{\theta\theta}$, and $\sigma_{zr} \approx 0$). Under these conditions, $\sigma_{zz} \approx 2 \lambda_r \mu_p \dot{\gamma}^2$. Drawing an analogy with Newton's constitutive relation, $\tau = \mu \dot{\gamma}$, we define an effective viscosity as $\mu_{\text{eff}} = 2 \mu_p \lambda_r \dot{\gamma}$. The average shear rate is estimated using the scaling developed in  \S\ref{sec:3-2-4}, $\dot{\gamma} = 4u_{\text{avg}}/e_2$, where $e_2$ denotes the distance between the filament tip and the channel wall (Figure~\ref{fig:5}(a)). The resulting value of $\mu_{\text{eff}}$ is subsequently substituted into equations  \eqref{eqn:27} and \eqref{eqn:28} to predict both the perturbation growth rate and the dominant wavelength. To validate the proposed formulation, we compare these predictions with measurements obtained from two representative fluids ($E_c=4.77$ and $E_c=10.87$) over a range of $Ca_1$ while maintaining constant $Ca_2$. The disturbance wavelength is measured as the spacing between neighbouring beads after each generation becomes clearly distinguishable. The corresponding growth rate is estimated by assuming exponential amplification of the disturbance, $d_f = d_0 \exp(\omega \Delta t)$, where $d_0$ and $d_f$ denote the initial and final bead diameters over the interval $\Delta t$. Since the growth rate evolves during bead development, the reported value represents the average growth rate over the complete formation process.

\begin{table}
\centering
\begin{tabular}{lcc ccccccccccc}
Fluids & \multicolumn{2}{c}{$Ca$} & Gen. & $e_2$ & $\dot{\gamma}$ & $\mu_{2,\dot{\gamma}}$ &  $\mu_{\text{eff.}}$ & $h_{\text{f}}$ & $\lambda$ & $\lambda_{\text{th}}$& $k$ &$\omega_{\text{exp.}}$ & $\omega_{\text{th}}$ \\

& & & & & & & & & & & & &\\

     & $Ca_2$ & $Ca_1$ &      &    $[\mu\text{m}]$ & $[s^{-1}]$ & $[\text{mPa-s}]$ & $[\text{mPa-s}]$ & $[\mu\text{m}]$ & $[\mu\text{m]}$&$[\mu\text{m]}$ & $[\mu\text{m}^{-1}]$ & $[s^{-1}]$ & $[s^{-1}]$\\

& & & & & & & & & & & & &\\
     
\midrule

\multirow{9}{*}{\begin{sideways} \textbf{F3} $(Ec = 4.77)$ \end{sideways}}& \multirow{3}{*}{0.0057} & \multirow{3}{*}{0.0161} & I  & 12  & 2037  & 10.3  & 3210  & 10  & 107& 46 & 0.059 & 93  & 92 \\
                    &                      &                      & II & 12 & 2037 & 10.3 & 3210  & 6.4 & 65& 31 & 0.097 & 117 & 156 \\
                    &                      &                      & III & 12 & 2037 & 10.3 & 3210  & 3.3 & 45&18 & 0.014& 152 & 175 \\ [5pt]

                     & \multirow{3}{*}{0.0057} & \multirow{3}{*}{0.0322} & I & 16 & 1458 & 12.5 & 2340 & 9.2 & 98& 39 & 0.064 & 122 & 114 \\
                    &                      &                      & II & 16 & 1458 & 12.5 & 2340 & 5.2 & 56& 24& 0.112 & 129 & 198 \\
                    &                      &                      & III & 16 & 1458 & 12.5 & 2340 & 2.3 & 40&12 & 0.157 & 160 & 181 \\[5pt]

                     & \multirow{3}{*}{0.0057} & \multirow{3}{*}{0.0483} & I  & 48  & 1435  & 12.6  & 2300  & 8.3  & 90&  36 & 0.070  & 126  & 123 \\
                    &                      &                      & II & 48  & 1435  & 12.6  & 2300  & 4.2  & 50 & 20& 0.126  & 150  & 206 \\ 
                    &                      &                      & III & 48  & 1435  & 12.6  & 2300  & 2.2  & 35 & 11 & 0.180  & 172  & 227 \\ [8pt]
                    
\multirow{9}{*}{\begin{sideways}\textbf{F5} $(Ec = 10.87)$ \end{sideways}} & \multirow{3}{*}{0.0057} & \multirow{3}{*}{0.0161} & I  & 48  & 509 & 14 & 4280 & 10.5 & 97& 57 & 0.065  & 31  & 38 \\
                    &                      &                      & II& 48  & 509 & 14 & 4280 & 7.5  & 75& 42 & 0.084  & 41  & 46 \\ 
                    &                      &                      & III & 48  & 509 & 14 & 4280 & 3.2  & 45& 23 & 0.140  & 55 & 58 \\ [5pt]

                   & \multirow{3}{*}{0.0057} & \multirow{3}{*}{0.0322} & I & 64 & 729 & 11 & 6060 & 9.5 & 80 &56 & 0.079 & 50 & 44 \\
                    &                      &                      & II & 64 & 729 & 11 & 6060 & 5.4 & 60 & 36 & 0.105 & 50 & 48\\
                    &                      &                      & III & 64 & 729 & 11 & 6060 & 2.6 & 35 & 20& 0.180 & 66 & 69  \\[5pt]

                    & \multirow{3}{*}{0.0057} & \multirow{3}{*}{0.0483} & I & 88 & 782 & 10.2 & 6490 & 8.2 & 70 &54 & 0.090 & 57 & 50 \\
                    &                      &                      & II & 88 & 782 & 10.2 & 6490 & 4.8 & 55 &36 & 0.114 & 59 & 50 \\
                    &                      &                      & III & 88 & 782 & 10.2 & 6490 & 2.2 & 29 &18 & 0.217 & 74 & 85  \\
\bottomrule
\end{tabular}
\caption{\justifying{Summary of wavelet characteristics during filament breakup for moderate and high elasticity fluids (i.e., $Ec = 4.77$, and $10.87$) at three different continuous phase capillary numbers ($Ca_1 = 0.0161, 0.0322,~\text{and}~0.0483$). The evolution of disturbance wavelength $\lambda$ and corresponding instability growth rate $\omega$ is reported for three distinct stages of the breakup process.}}
\label{tab:table2}
\end{table}

The results are summarised in Table~\ref{tab:table2}. For both fluids, the effective viscosity increases with increasing $Ca_1$, primarily because the increase in mean velocity raises the interfacial shear rate. A further increase in elasticity (or $E_c$) produces an additional increase in $\mu_{\text{eff}}$ through the larger relaxation time. Although highly elastic fluids exhibit greater droplet lift, resulting in a larger wall distance $e_2$ and hence a slight reduction in the average shear rate, this effect is outweighed by the combined increase in flow velocity and polymer relaxation time. Consequently, $\mu_{\text{eff}}$ increases monotonically with both $Ca_1$ and $E_c$. 

The measured instability characteristics also evolve systematically through successive generations. As the filament progressively thins, its diameter $h_f$ decreases, leading to a reduction in the dominant wavelength predicted by equation~\eqref{eqn:28}. This trend is consistent with the experimental observations, where each new generation develops within progressively thinner connecting threads. In contrast, the disturbance growth rate increases from the first to the third generation. This acceleration arises from several concurrent mechanisms. First, continued filament thinning is accompanied by an increase in the axial flow velocity (Figure~\ref{fig:6}(b)), which enhances disturbance amplification. Second, first-generation beads develop on relatively thick filaments and therefore require larger fluid volumes before becoming distinguishable, whereas later generations form on much thinner threads and attain observable amplitudes more rapidly. Finally, capillary-driven transport continuously transfers liquid from the interconnecting threads into adjacent beads, allowing the smaller higher-generation beads to grow faster than the larger first-generation structures. The comparison presented in Table~\ref{tab:table2} demonstrates that incorporating the effective viscosity into the classical Rayleigh$-$Plateau formulation yields predictions of both the instability wavelength and the disturbance growth rate that agree with the experimental measurements to the correct order of magnitude. Although viscoelastic filament breakup differs fundamentally from the classical Rayleigh--Plateau instability because of the additional contribution of elastic stresses, the present results show that its essential instability characteristics can nevertheless be captured within an appropriately modified RPI framework based on an effective viscoelastic viscosity. This provides a physically consistent bridge between classical capillary instability theory and the breakup dynamics of confined viscoelastic filaments.

\subsection{Secondary droplet distribution and uniformity}
\label{sec:3-5}

The final stage of filament breakup determines the size distribution and uniformity of the secondary droplets generated from the viscoelastic filament. Since the breakup proceeds through successive generations of capillary instability, both the number and the size of the secondary droplets depend on the filament length established during the preceding stretching stage. As shown in \S\ref{sec:3-2-4}, increasing the capillary numbers and the elastocapillary number increases the attainable filament length before breakup. The present section therefore examines how these parameters influence the statistics of the resulting secondary droplets.

\begin{figure}
  \centerline{\includegraphics[width=1.0\columnwidth]{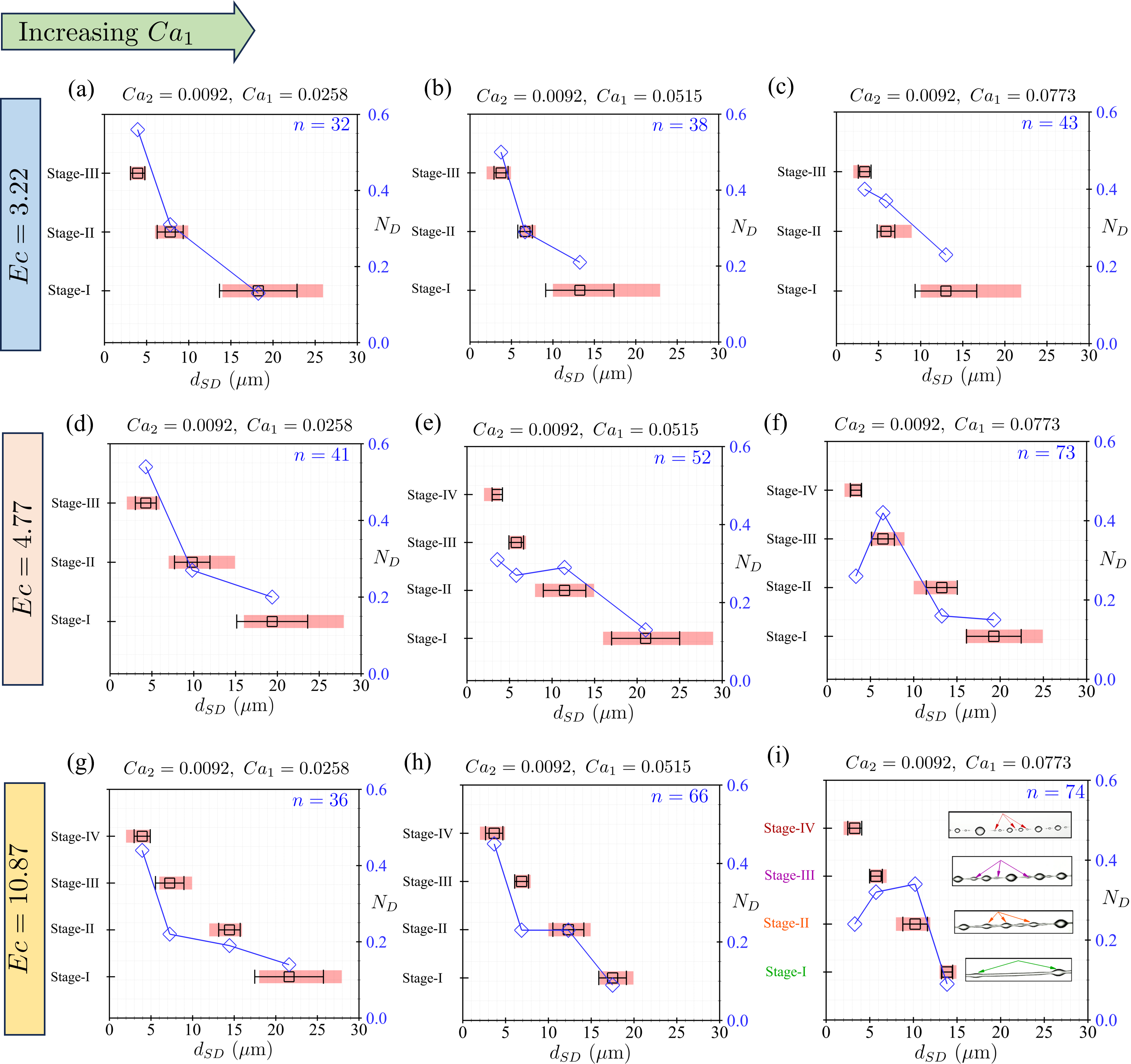}}
   \medskip
  \caption{\justifying{Distribution of secondary droplet sizes for different elastocapillary numbers ($Ec$), showing the dependence on continuous-phase capillary number ($Ca_1$) while the dispersed-phase capillary number ($Ca_2$) is held constant. The results highlight the interplay between capillary, viscous, and elastic effects in satellite droplet formation.
(a–c) Evolution of the mean diameter of satellite droplets, $d_{SD}$, at different stages of filament instability with increasing $Ca_1$ for $Ec = 3.22$. The number density of secondary droplets, $N_{SD}$, increases at later stages. (d–f) and (g–h) show the corresponding distributions for $Ec = 4.77$ and $Ec = 10.77$, respectively, exhibiting similar trends with a progressive shift in $d_{SD}$ and an enhanced $N_{SD}$ at later stages of the instability. Moreover, the polydispersity decreases with increasing $Ca_1$ and $Ec$, indicating a more uniform satellite droplet size distribution at higher continuous-phase capillary numbers.}}
\label{fig:10}
\end{figure}

Figure~\ref{fig:10} summarises the generation-wise evolution of the secondary droplet population for three representative elasticities ($E_c=3.22, 4.77$ and $10.87$) while maintaining a fixed dispersed-phase capillary number ($Ca_2=0.0092$). The continuous-phase capillary number increases from left to right. For each condition, the mean secondary droplet diameter ($d_{SD}$), the total number of secondary droplets ($n$), and the generation-wise number density $N_D=n_{i}/n$, are evaluated, where $n_i$ denotes the number of droplets produced in generation $i$. Representative experimental images illustrating the successive breakup process are included in Figure~\ref{fig:10}(i). For the weakly elastic fluid ($E_c=3.22$), filament breakup is completed within three generations. Increasing $Ca_1$ produces progressively thinner and longer filaments, leading to a reduction in the mean secondary droplet diameter together with a simultaneous increase in the total droplet population. The droplet size distribution narrows from approximately $4-18\mu$m at $Ca_1=0.0258$ to $4-13\mu$m at $Ca_1=0.0773$, while the total droplet number increases by approximately $34\%$. At the same time, the contribution of the third generation decreases relative to the earlier generations, indicating that stronger stretching redistributes breakup events towards the initial generations rather than promoting repeated fragmentation. Consequently, increasing $Ca_1$ simultaneously increases droplet yield and reduces polydispersity.

The influence of elasticity becomes progressively more pronounced as $E_c$ increases. Raising the elasticity from $E_c=3.22$ to $E_c=4.77$ increases the total droplet number by approximately $28\%$ even at the lowest capillary number, while only marginally altering the overall droplet-size distribution. At larger $Ca_1$, however, the longer filaments generated by the more elastic fluid undergo an additional generation of breakup, resulting in complete fragmentation only after the fourth generation. For the highly elastic fluid ($E_c=10.87$), four generations of breakup are observed even at the lowest capillary number investigated. As $Ca_1$ increases, the mean secondary droplet diameter decreases steadily while the uncertainty associated with each generation also decreases, demonstrating a significant improvement in droplet uniformity. At $Ca_1=0.0773$, the droplet diameter is largely confined to the range $4-14~\mu$m, while the total number of droplets is approximately $72\%$ larger than that obtained for the weakly elastic fluid. These observations demonstrate that elasticity primarily increases the breakup hierarchy by delaying capillary rupture, thereby producing a larger number of smaller and more uniform droplets.

\begin{figure}
  \centerline{\includegraphics[width=0.85\columnwidth]{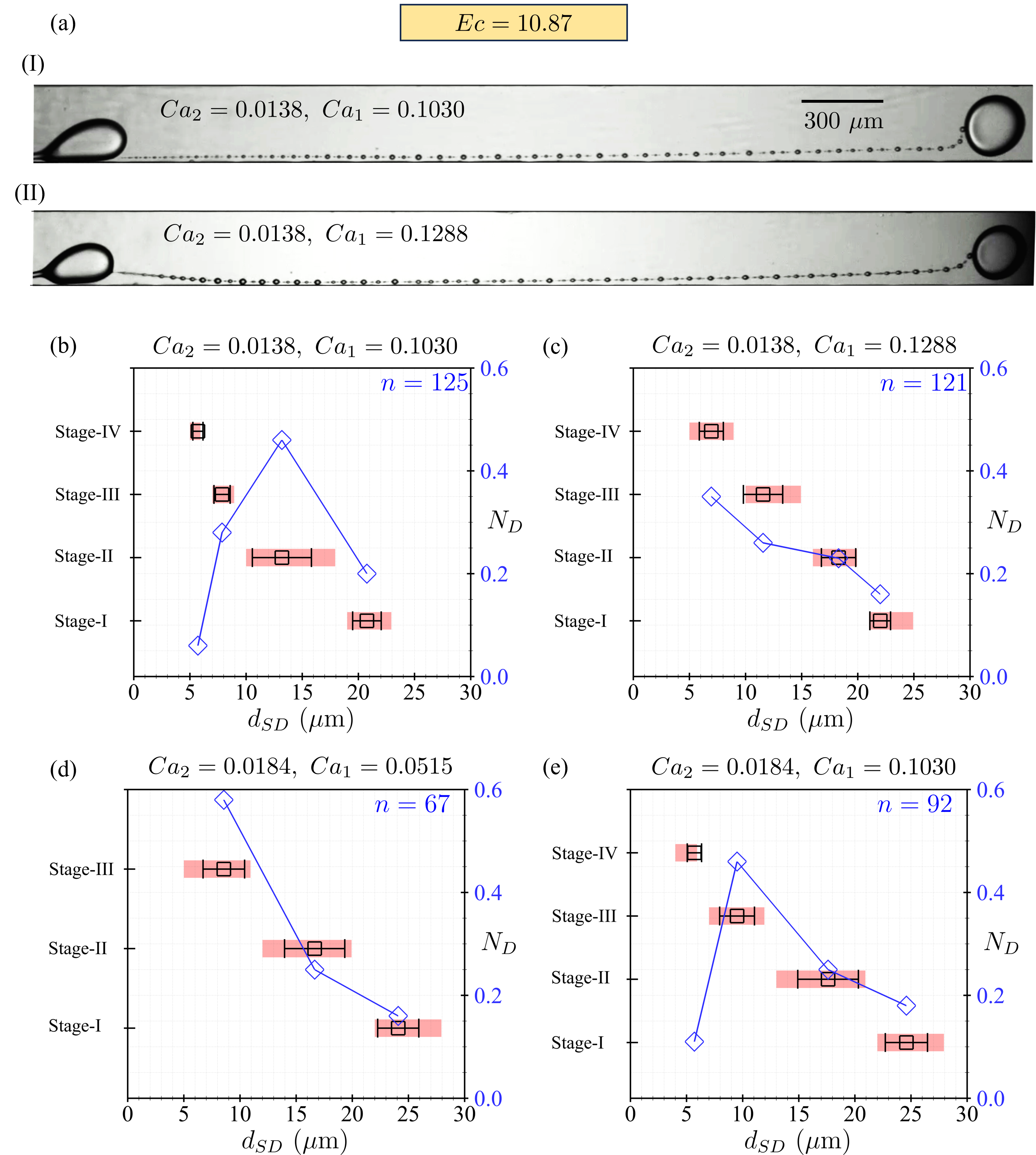}}
   \medskip
  \caption{\justifying{ Effect of capillary numbers on satellite droplet formation and uniformity. (a) Representative snapshots at fixed $Ca_2 = 0.0138$ and varying $Ca_1$, illustrating the formation of a primary droplet followed by a train of satellite droplets. Scale bar: $300~\mu$m. 
(b--e) Quantitative characterization of satellite droplet breakup across different stages (Stage-I to Stage-IV) for varying $(Ca_1, Ca_2)$. The mean satellite droplet diameter $d_{SD}$ is shown along the horizontal axis, while the normalized number of satellite droplets $N_D$ is shown on the right axis. Error bars indicate the spread in $d_{SD}$, and $n$ denotes the total number of secondary droplets generated in each case. 
At fixed $Ca_2 = 0.0138$ (b--c), increasing $Ca_1$ shifts the distribution toward smaller $d_{SD}$ with reduced dispersion, indicating improved uniformity of satellite droplets at later breakup stages. In contrast, at higher $Ca_2 = 0.0184$ (d--e), the distributions broaden and shift toward larger $d_{SD}$, reflecting increased variability and reduced uniformity due to stronger hydrodynamic effects. Across all cases, $N_D$ exhibits a non-monotonic variation with $d_{SD}$, with peak satellite production occurring at intermediate stages corresponding to active interfacial necking. These results demonstrate that lower $Ca_2$ combined with moderate $Ca_1$ promotes more uniform satellite droplet formation, whereas increasing $Ca_2$ leads to a transition toward less controlled, hydrodynamically dominated breakup.}}
\label{fig:11}
\end{figure}

The above observations are consistent with the filament-length scaling developed in \S\ref{sec:3-2-4}. According to equation~\eqref{eqn:23},

\begin{equation}
{l}_f \sim Ec (\beta \,Ca_1+Ca_2)^2
\label{eqn:29}
\end{equation}
indicating that increasing either elasticity or the imposed shear extends the lifetime and length of the filament before breakup. Longer filaments provide a larger spatial extent over which capillary disturbances can develop, allowing successive generations of beads to emerge before complete rupture. Consequently, increasing $E_c$, $Ca_1$, or $Ca_2$ promotes repeated fragmentation, leading simultaneously to a larger droplet population and reduced droplet-size variability.

Motivated by this scaling, we next examine breakup under larger capillary numbers, where substantially longer filaments are produced. Figure~\ref{fig:11} examines the secondary droplet distribution under larger values of $Ca_1$ and $Ca_2$. Increasing both capillary numbers initially increases the total droplet yield while simultaneously narrowing the droplet-size distribution. For example, at $Ca_2=0.0138$ and $Ca_1=0.1030$, approximately $75\%$ of the droplets lie within the relatively narrow diameter range of $7-13~\mu$m. Under these conditions the breakup process remains highly repeatable, producing a nearly monodisperse droplet population. Further increasing the capillary numbers alters the filament configuration near the channel wall. Part of the filament remains attached close to the wall before breakup, giving rise to a small number of larger droplets while preserving a relatively narrow dominant size distribution. At still larger $Ca_2$, the stronger stretching produces larger primary filaments and shifts the secondary droplet distribution towards larger diameters ($8-24~\mu$m), thereby increasing polydispersity despite the increase in droplet production. Thus, although increasing the capillary numbers generally favours higher droplet yield, excessively large $Ca_2$ eventually compromises droplet uniformity through wall-mediated breakup. The corresponding droplet-size distributions are summarised quantitatively in Table~\ref{tab:table3}. The results demonstrate that specific combinations of $Ca_1$ and $Ca_2$ can be selected to maximise the production of droplets within prescribed diameter ranges. For example, Case II yields the largest proportion of droplets near $7~\mu$m, whereas Case I is more suitable for producing droplets near $12~\mu$m. More generally, the cumulative distributions reported in Table~\ref{tab:table3} provide a practical design map for selecting operating conditions to generate droplets within any desired size interval.

These results establish that the combined action of viscous stretching and viscoelastic stresses not only governs filament formation and breakup but also provides a robust mechanism for tuning the size, yield, and uniformity of micron-sized droplets. Such control is particularly attractive for microfluidic applications involving encapsulation, emulsification, single-cell analysis, and high-throughput screening, where both droplet monodispersity and production rate are critical.

\begin{table}
\centering
\begin{tabular}{ccccccccc}
\multicolumn{9}{c}{$Ec = 10.87$ (\textbf{F5})} \\[5pt]
Case &\multicolumn{2}{c}{Capillary numbers} &   & \multicolumn{5}{c}{Percentage of secondary droplet} \\[2pt]
\cmidrule(lr){2-3} \cmidrule(lr){4-8}
 &$Ca_2$ & $Ca_1$ & $n$ &  $7 \mp 2~\mu$m & $12 \mp 2~\mu$m & $17 \mp 2~\mu$m & $22 \mp 2~\mu$m & $27 \mp 2~\mu$m \\[5pt]
\midrule

I &0.0138 & 0.1030 & 125 & 33.6 & 32 & 17.6 & 16.8 & -- \\[2pt]
II &0.0138 & 0.1288 & 121 & 34.7 & 25.6 & 14.9 & 24 & 0.8 \\[2pt]
III &0.0184 & 0.0515 & 67 & 37.3 & 28.3 & 12 & 17.9 & 4.5 \\[2pt]
IV &0.0184 & 0.1030 & 92 & 34.8 & 26 & 12 & 18.5 & 8.7 \\

\bottomrule
\end{tabular}
\caption{\justifying{Size distribution of satellite droplets for different $(Ca_1, Ca_2)$, showing the percentage of droplets within specified diameter ranges and total count $n$. Lower $Ca_2 = 0.0138$ yields a narrower distribution with a higher fraction in smaller size ranges (5--12~$\mu$m), indicating better uniformity, whereas higher $Ca_2 = 0.0184$ leads to a broader distributions. Increasing $Ca_1$ shifts the distribution toward smaller droplets but does not fully recover uniformity at higher $Ca_2$.}}
\label{tab:table3}
\end{table}

\section{Conclusions}
\label{sec:4}

We have experimentally investigated the generation, stretching and breakup of confined viscoelastic filaments formed by a shear-thinning dispersed phase driven by a Newtonian continuous phase in a microfluidic co-flow junction. By systematically varying the capillary numbers of the two phases and the fluid relaxation time, we establish the physical mechanisms governing flow-regime transitions, filament evolution, capillary instability and secondary droplet formation. Through high-speed imaging, confocal microscopy, particle tracking and scaling analysis, we develop a unified physical framework for filament-mediated breakup in confined viscoelastic flows. The experiments demonstrate that the onset of droplet generation is governed primarily by the hydrodynamic competition between the two phases and is largely independent of fluid elasticity. Over the range of conditions investigated, the transition from stable co-flow to droplet formation collapses onto the universal criterion ($Ca_1 \approx 2.3~Ca_{2,\dot{\gamma}}$). Once a filament is formed, however, elasticity fundamentally alters its subsequent evolution by suppressing capillary thinning, prolonging filament lifetime and promoting the transition from dripping to jetting. Thus, elasticity plays only a minor role in initiating breakup but becomes the dominant mechanism controlling its evolution.

The principal finding of this work is that confined viscoelastic filament dynamics are governed by the coupled action of polymer elasticity and wall-induced interfacial shear. As the primary droplet is advected downstream, it stretches the connecting filament while simultaneously lifting it away from the channel wall, producing an inclined filament subjected to a highly non-uniform shear field. Direct particle-tracking measurements reveal pronounced spatial variations in the interfacial velocity along the filament, providing direct experimental evidence that wall-induced shear actively regulates filament stretching, thinning and eventual breakup. This coupling between confinement-induced shear and elastic stresses represents the defining physical mechanism of confined viscoelastic filament dynamics and distinguishes them fundamentally from the classical elastocapillary thinning of freely suspended liquid bridges. Guided by these observations, we derive physically transparent scaling relations from balances between capillary pressure, viscous shear and elastic tensile stresses. A single scaling framework successfully predicts the principal geometric characteristics of the breakup process, including the primary droplet diameter, the critical filament thickness at the onset of instability, the maximum filament length in the dripping regime and the critical thread length in the jetting regime. Unlike conventional elastocapillary descriptions that consider only extensional deformation, the present framework explicitly incorporates the interfacial shear imposed by the surrounding continuous phase, thereby extending classical filament-thinning theory to confined microfluidic flows.

The experiments further reveal that filament instability is spatially organised rather than random. The first beads-on-a-string perturbation consistently originates at the location of maximum wall-induced shear, after which successive generations of beads emerge through repeated instability of progressively thinner connecting threads. Although viscoelastic breakup differs fundamentally from the classical Rayleigh–Plateau instability owing to the presence of elastic stresses, the dominant instability characteristics, including the disturbance wavelength and growth rate, can nevertheless be captured within a modified Rayleigh–Plateau framework by introducing an effective viscosity based on the polymeric normal stress. Finally, we show that elasticity provides a robust mechanism for controlling secondary droplet production. Increasing elasticity and continuous-phase shear produces longer filaments that generate larger numbers of secondary droplets with reduced size variability, while an operating window below the onset of jetting enables reproducible generation of highly uniform micron-sized droplets. The present study establishes a coherent physical picture of confined viscoelastic breakup in which capillary pressure, wall-induced shear and polymer elasticity act together to determine filament stretching, instability and fragmentation. More broadly, the results demonstrate that confinement fundamentally modifies elastocapillary breakup by introducing wall-induced shear as an additional control parameter that couples directly with elastic stress development. The experimentally validated scaling laws and mechanistic insights presented here provide a predictive framework for confined multiphase flows involving complex fluids and should facilitate the rational design of microfluidic systems for controlled droplet generation and related applications.


\appendix

\renewcommand{\thefigure}{A\arabic{figure}}
\setcounter{figure}{0}

\section{Fabrication of the microfluidic device}\label{appA}
The microchannel device is fabricated in PDMS using standard photolithography and soft lithography techniques, as described below. A photomask is designed using AutoCAD 2015 and printed by JD Photo Data (UK). The silicon wafer used for the photolithography process is cleaned sequentially using RCA1, RCA2, and an HF dip, followed by a DI water rinse, and then placed in an oven at $120^{\circ}\text{C}$ for 2 min to remove moisture. The photoresist SU8 2075 (MicroChem Corp., Newton, USA) is spin-coated onto the wafer at 2100 rpm for 30 s with an acceleration of 300~\text{rpm~s} $^{-1}$, followed by soft baking at $65^{\circ}\text{C}$ for 5 min and $95^{\circ}\text{C}$ for 10 min. The photoresist is then exposed to UV light through the photomask for 30 s.  Post-exposure baking is carried out at $65^{\circ}\text{C}$ for 2 min followed by $95^{\circ}\text{C}$ for 8 min. Subsequently, the UV-exposed wafer is developed to obtain the silicon master with the SU8 pattern on top. The developed wafer is further baked in an oven at $100^{\circ}\text{C}$ for 30 min to improve the adhesion between the photoresist and the wafer. The dimensions of the SU8 pattern are verified using the Veeco NT-1100 optical surface profiler and a confocal microscope, which scan the surfaces using an incident light source and measure the emissive, reflective, or refractive light to obtain information about the surface topography. The PDMS monomer and curing agent (Sylgard 184, Dow Corning, USA) are mixed in a 10:1 ratio by weight, and the mixture is then degassed (in a desiccator) before being poured onto the fabricated silicon master. After pouring, the PDMS is cured in a vacuum oven at $75^{\circ}\text{C}$ for 1 h. The cured PDMS layer containing the channel structure is then peeled off from the silicon master and cut to the required size. Fluidic access holes for the inlet, outlet, and pressure taps are punched using a 1.5 mm biopsy punch (Med Morphosis LLP, Pondicherry, India). Subsequently, the PDMS layer containing the microchannel structure is bonded to a glass slide using an oxygen plasma bonder from Harrick Plasma. The fabricated device consists of an expanded channel section with a width of 300 $\mu \text{m}$ and a depth of 100 $\mu \text{m}$.

\section{Shear-rate dependence of viscosity}\label{appB}
\begin{figure}
  \centerline{\includegraphics[width=0.4\columnwidth]{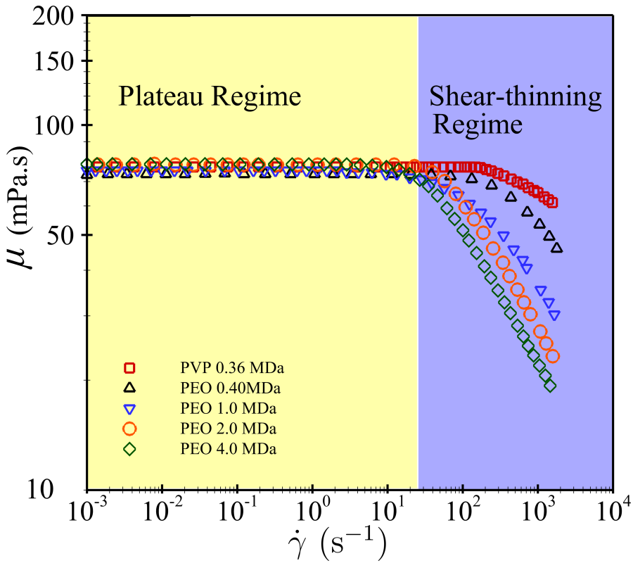}}

  \caption{Viscosity variation with shear rate}
\label{fig:B1}
\end{figure}

The variation of viscosities ($\mu$) with shear rate ($\dot{\gamma}$) for different polymeric fluids : PEO (0.4 MDa, 2.4$\%$), PEO (1 MDa, 1.7 $\%$), PEO (2 MDa, 1.15$\%$), PEO (4 MDa, 0.97 $\%$), and PVP (0.36 MDa, 8.5 $\%$) are presented in Figure~\ref{fig:B1}. At low shear rates ($\dot{\gamma} \leq 30~s^{-1}$), shear-thinning effects are negligible for all the fluids. However, at higher shear rates, pronounced shear-thinning behaviour is observed. The extent of shear-thinning is more significant in high–molecular weight PEO solutions.

\section{Demarcation of primary flow regimes}\label{appC}
\begin{figure}
  \centerline{\includegraphics[width=0.4\columnwidth]{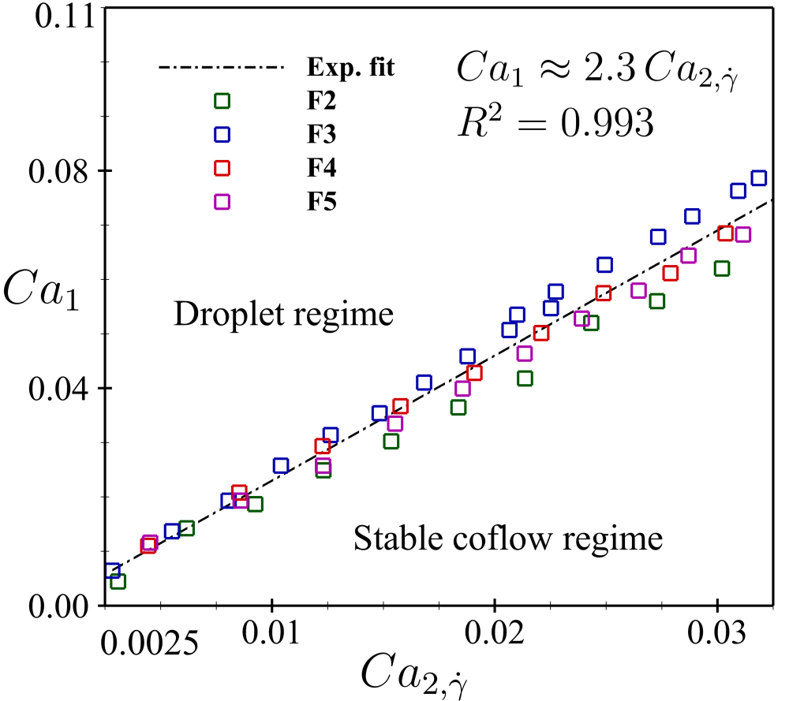}}
  \caption{\justifying{A regime map ($Ca_{2,\dot{\gamma}}$ versus $Ca_1$) showing the transition from stable coflow to droplet regimes. The demarcation line (dash dot) separating the two regimes collapses onto a single boundary for fluids with different elasticity ($Ec$), indicating that the transition is largely independent of $Ec$.}}
\label{fig:C1}
\end{figure}

When the immiscible parallel flow of a viscoelastic fluid ($P_2$) and a Newtonian fluid ($P_1$) transitions from stable coflow to droplet forming regimes, a variety of flow configurations, namely squeezing, dripping, and jetting, interconnected by thin filaments, emerge. Accordingly, the flow behaviour can be broadly classified into two primary regimes: stable coflow and droplet regimes, the latter encompassing all droplet formation modes. Figures~\ref{fig:2}(b)–(e) present the regime maps for fluids with different elastic properties ($Ec$), where the dash dot line denotes the locus of transition points ($Ca_{2,\dot{\gamma}}, Ca_1$) separating stable coflow from droplet formation. To examine the influence of elasticity, all transition points are superposed in Figure~\ref{fig:C1}. Remarkably, the transition points for all elastic fluids collapse onto a narrow band, indicating that the transition from stable coflow to droplet formation is largely independent of $Ec$. Furthermore, the transition boundary is well described by the linear relation $Ca_1 \approx 2.3,Ca_{2,\dot{\gamma}}$.

\section{Empirical relation for droplet neck width}\label{appD}
\begin{figure}
  \centerline{\includegraphics[width=0.4\columnwidth]{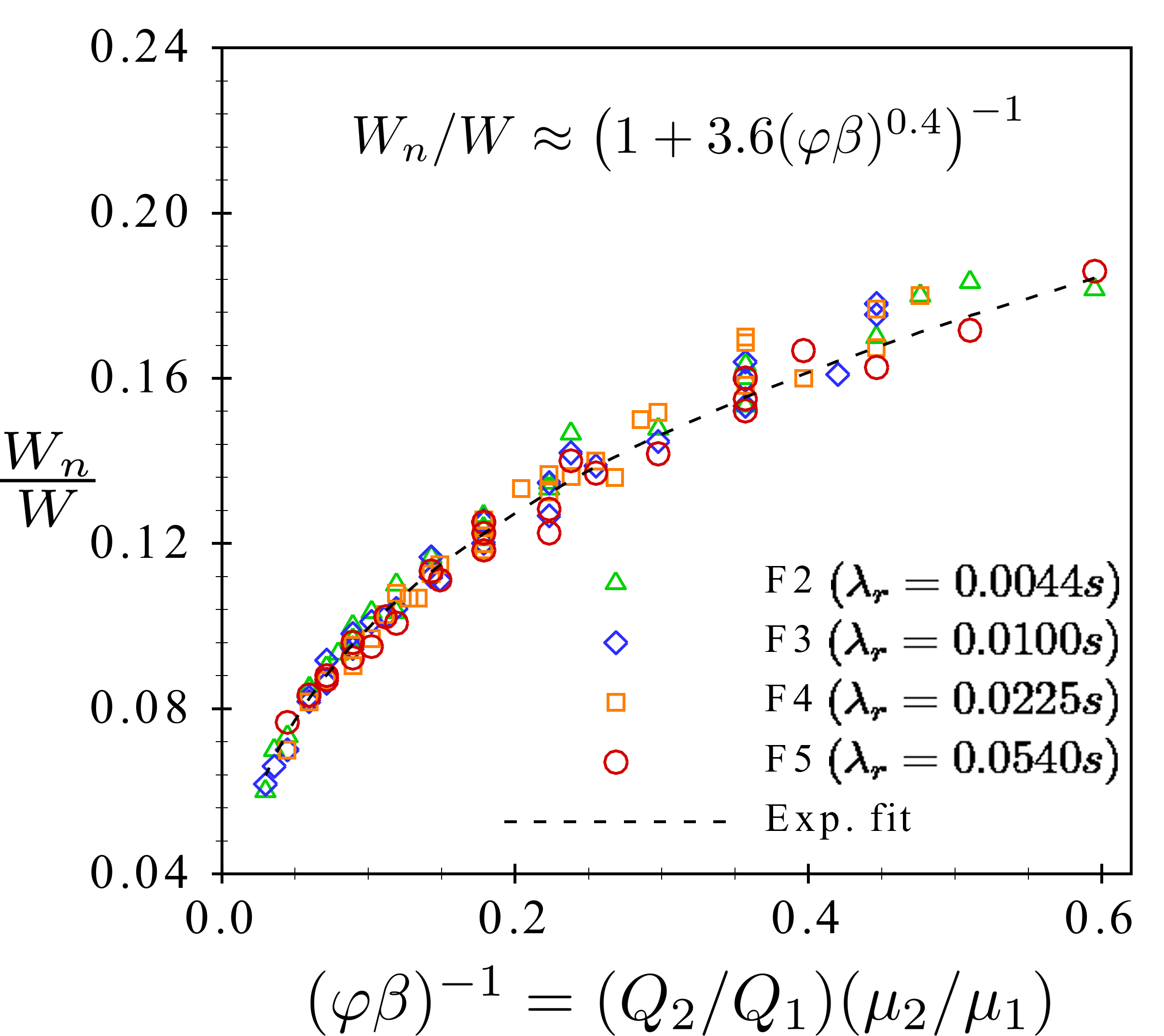}}
  \caption{\justifying{Evolution of the dimensionless droplet neck width ($W_n/W$) in the squeezing and dripping regimes as a function of $(Q_r m)^{-1} = (Q_2/Q_1)(\mu_2/\mu_1)$ for fluids with different relaxation times ($\lambda_r$). The dashed line represents the experimental fit, $W_n/W \approx \left(1 + 3.6(\varphi\beta)^{0.4}\right)^{-1}$.}}
\label{fig:D1}
\end{figure}

During droplet generation in the squeezing and dripping regimes, the dispersed phase ($P_2$) penetrates into the main channel and progressively expands into a bulb shaped interface. As the growing droplet increasingly obstructs the channel, the continuous phase ($P_1$) is forced through the narrow gaps surrounding the droplet, producing enhanced viscous shear and a pressure buildup near the junction. Consequently, a neck forms between the emerging droplet and the upstream fluid reservoir (see Figure~\ref{fig:3}(a)); the corresponding neck width is denoted by $W_n$. Experimental observations show that the dimensionless neck width ($W_n/W$) primarily depends on the flow rate ratio ($\varphi$). Earlier studies \citep{CubudTMason2008,Hazra2019} demonstrated that, for stable coflow, the width ratio ($W_2/W$) scales with both the flow rate ratio ($\varphi$) and the viscosity ratio ($\beta$), and proposed the empirical relation, $W_2/W \approx (1+a(\varphi \beta)^b)^{-1}$, where $a$ and $b$ are fitting constants. Motivated by these observations, we plot $W_n/W$ as a function of $(\varphi \beta)^{-1}$ for fluids with different relaxation times ($\lambda_r$), as shown in Figure~\ref{fig:D1}. Remarkably, the data for different $\lambda_r$ collapse onto a single master curve, indicating that the neck width is essentially independent of polymer relaxation time. The experimental data are well described by the correlation $W_n/W \approx (1+3.6(\varphi \beta)^{0.4})^{-1}$.

\section{Elastocapillary thinning and relaxation time}
\label{appE}
\begin{figure}
  \centering{\includegraphics[width=0.95\columnwidth]{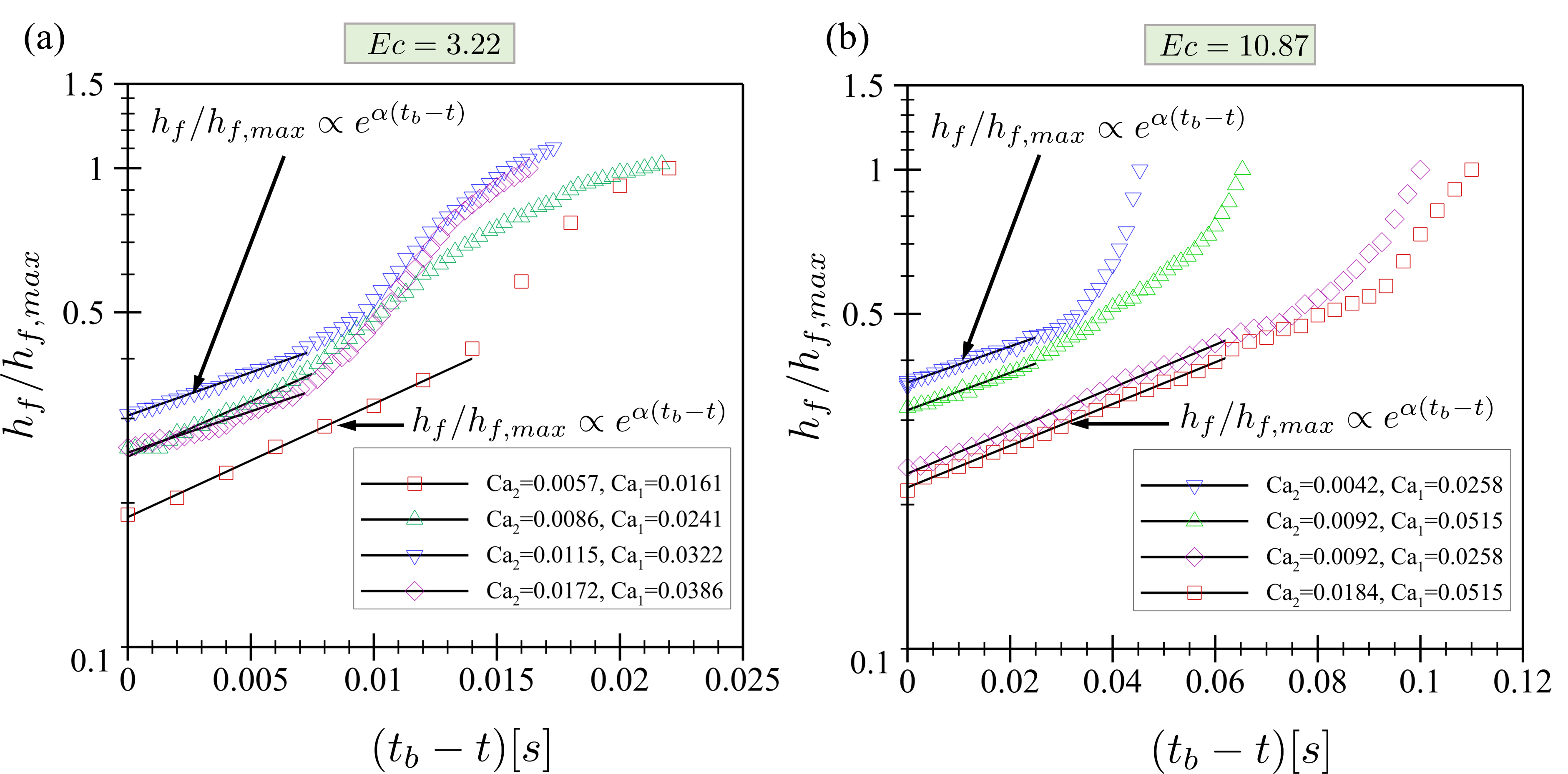}}
   \medskip
  \caption{\justifying{Variation of the normalized filament width, $h_f/h_{f,\max}$, as a function of the time remaining to breakup, $(t_b-t)$, for (a) a moderately elastic fluid ($Ec = 3.22$) and (b) a highly elastic fluid ($Ec = 10.87$) under four different combinations of $Ca_1$ and $Ca_2$. The symbols represent the experimental measurements, while the solid lines denote the corresponding exponential fits.}}
\label{fig:E1}
\end{figure}

We investigate the influence of confinement and imposed shear on the pinch-off dynamics of viscoelastic fluids by analysing the evolution of the filament width during droplet breakup. Figure~\ref{fig:E1} presents the variation of the dimensionless filament width, $h_f/h_{f,\max}$, as a function of the relative breakup time, $(t_b-t)$, on semi-logarithmic scales for a moderately elastic fluid (PEO 1MDa, 1.7 wt.$\%$ aqueous solution, $Ec=3.22$) under four different combinations of the continuous- and dispersed-phase capillary numbers, $Ca_1$ and $Ca_2$, respectively.

At the onset of necking, the filament width reaches its maximum value, $h_{f,\max}$ ($=W_n$, Figure~\ref{fig:1}). As the breakup process proceeds, the filament continuously thins until it becomes unstable. The instant at which the filament loses its stability and a bead-on-a-string structure first appears is defined as the breakup time, $t_b$. To characterise the thinning dynamics, the filament width is plotted against the relative time before breakup, $(t_b-t)$.

Particular attention is paid to the late-stage thinning, corresponding to the elastocapillary regime ($h_f/h_{f,\max}< 0.5$). As shown in Figure~\ref{fig:E1}, the filament exhibits a clear exponential decrease in width in this regime for all investigated flow conditions. Such exponential thinning is a characteristic feature of viscoelastic filament breakup and has been extensively studied in extensional rheometry, where interfacial shear is negligible \citep{OliveiraYeh2006,ArdekaniSharma2010,ClasenEggers2006}. In contrast, the corresponding dynamics in confined shear flows have received relatively little attention \citep{Husny2006,Steinhaus2007}. Previous studies \citep{EntovHinch1997,Anna2001,ClasenEggers2006} have demonstrated that this behaviour arises from a local balance between the capillary pressure driving filament thinning and the elastic stresses generated by stretched polymer molecules. Notably, \cite{ClasenEggers2006} developed a one-dimensional theoretical model and validated it against the experimental observations of \cite{Anna2001}. Their analysis showed that, for $t>\lambda_r$, the filament radius decreases exponentially according to

\begin{equation}
h_f(t)/h_{f,max} = \left(\frac{Gh_{f,max}}{2\gamma}\right)^{1/3}~e^{-t/3\lambda_r},
 \label{eqn:E1}   
\end{equation}
where $G(=\eta_p/\lambda_r)$ is the elastic modulus, $\eta_p=(\eta_0-\eta_s)$ is the polymeric contribution to the solution viscosity, and $\eta_0$ and $\eta_s$ denote the zero-shear viscosity of the polymer solution and the solvent, respectively.

Motivated by this theoretical framework, we determine the polymer relaxation time, $\lambda_r$, by fitting Eqn.~(\ref{eqn:E1}) to the experimentally measured filament-thinning data in the elastocapillary regime ($t>\lambda_r$). Following the methodology adopted in previous studies \citep{Husny2006,Steinhaus2007}, the relaxation time is extracted for each combination of $Ca_1$ and $Ca_2$, enabling us to quantify how confinement and imposed shear influence the extensional relaxation dynamics of the viscoelastic filament.

Figure~\ref{fig:E1}(a) shows that, for the moderately elastic fluid ($Ec = 3.22$), the filament-thinning data obtained under different combinations of $Ca_1$ and $Ca_2$ exhibit an exponential decay of the form $h_f/h_{f,\max} \propto e^{\alpha t}$ in the elastocapillary regime, shown as solid lines. Although four different combinations of $Ca_1$ and $Ca_2$ were investigated, the first two cases (represented by the square and triangle symbols) correspond to the same capillary number ratio, $Ca_r = Ca_2/Ca_1 = 0.35$, whereas the remaining two cases (represented by the inverted triangle and diamond symbols) have a significantly larger ratio of $Ca_r \approx 0.46$. The exponential fits yield decay coefficients of $\alpha = 42~\mathrm{s}^{-1}$ and $56~\mathrm{s}^{-1}$ for $Ca_r = 0.35$ and $Ca_r \approx 0.46$, respectively. Accordingly, the polymer relaxation time, obtained from $\alpha = 1/(3\lambda_r)$, decreases from $0.008~\mathrm{s}$ to $0.006~\mathrm{s}$ as the capillary number ratio increases from $0.35$ to approximately $0.46$. Figure~\ref{fig:E1}(b) presents the filament-thinning characteristics of the highly elastic fluid. The same qualitative behaviour is observed, with exponential thinning in the elastocapillary regime and a relaxation time that varies with the capillary number ratio. For the first two capillary-number combinations (represented by the inverted triangle and triangle symbols), corresponding to $Ca_r \approx 0.17$, the exponential fit yields a decay coefficient of $\alpha = 8.8~\mathrm{s}^{-1}$. For the remaining two combinations (represented by the square and diamond symbols), corresponding to $Ca_r \approx 0.36$, the decay coefficient increases to $\alpha = 10.2~\mathrm{s}^{-1}$. Accordingly, the estimated polymer relaxation time is inversely related to the capillary-number ratio, increasing from $\lambda_r = 0.033~\mathrm{s}$ at $Ca_r \approx 0.36$ to $\lambda_r = 0.038~\mathrm{s}$ at $Ca_r \approx 0.17$. Thus, a lower capillary-number ratio results in a larger apparent relaxation time.

The physical implication of a lower $Ca_r$ is that, for a fixed dispersed-phase capillary number ($Ca_2$), the continuous-phase capillary number ($Ca_1$) is higher, resulting in a stronger shear imposed by the continuous phase. Thus, the results indicate that increasing the continuous-phase shear leads to an increase in the apparent polymer relaxation time, $\lambda_r$, consistent with previous experimental observations \citep{Husny2006}.Thus, the present results indicate that the apparent polymer relaxation time, $\lambda_r$, extracted from the filament-thinning data using the correlation of Clasen \textit{et al.} \citep{ClasenEggers2006}, is not solely an intrinsic material property of the polymer solution. Rather, under the present confined co-flow conditions, it also depends on the capillary number ratio, $Ca_r$, indicating that the imposed shear modifies the apparent elastocapillary thinning dynamics.

\begin{table}
  \begin{center}
\def~{\hphantom{0}}
  \begin{tabular}{lccccc}
  & \multicolumn{5}{c}{Polymer relaxation times $\lambda_r~[s]$}\\[3pt]
        & PEO 0.4MDa    & PEO 1.0MDa &   PEO 2.0MDa &  PEO 4.0MDa  &  PVP 0.36MDa    \\[5pt]
\hline
       & (2.5 wt.$\%$)   & (1.7 wt.$\%$) &  (1.15 wt.$\%$) & (0.95 wt.$\%$) & (8.5 wt.$\%$)  \\[3pt]
      From Literature & 0.0044 & 0.010 & 0.0225 & 0.0540 & 0.0013 \\
       From Experiment & 0.0045 & 0.00805 & 0.0193 & 0.0433 & 0.0011 \\
       & ($\pm$ 0.0003) & ($\pm$ 0.001) & ($\pm$ 0.003) & ($\pm$ 0.009) & ($\pm$ 0.0004) \\
\hline      
  \end{tabular}
  \caption{\justifying{Comparison of the longest polymer relaxation times for the viscoelastic fluids investigated in the present study. The relaxation times estimated from filament-thinning experiments are compared with corresponding literature values.}}
  \label{tab:table4}
  \end{center}
\end{table}

Indeed, the present results indicate that the estimated polymer relaxation time varies with the capillary numbers. Therefore, to facilitate comparison with values reported in the literature \citep{Ebagninin2009,Arnolds2010,Naillon2019}, Table~\ref{tab:table4} presents a representative relaxation time for each fluid, obtained by averaging the relaxation times estimated over all capillary-number combinations investigated in the present study. The Table~\ref{tab:table4} also reports the upper and lower bounds of the estimated relaxation time. The upper bound corresponds to the longest relaxation time extracted from the filament-thinning data, which shows good agreement with the values reported in the literature\citep{Ebagninin2009,Arnolds2010,Naillon2019}.

\counterwithin*{equation}{section}
\renewcommand\theequation{\thesection\arabic{equation}}

\bibliographystyle{jfm}
\bibliography{jfm}

\clearpage
\section*{Supplementary material}
\addcontentsline{toc}{section}{Supplementary material}

\setcounter{section}{0}
\setcounter{figure}{0}
\setcounter{table}{0}
\setcounter{equation}{0}

\renewcommand{\thefigure}{S\arabic{figure}}
\renewcommand{\thetable}{S\arabic{table}}
\renewcommand{\theequation}{E\arabic{equation}}

\newcommand{\suppsection}[1]{%
  \refstepcounter{section}%
  \section*{S\arabic{section}. #1}%
  \addcontentsline{toc}{section}{S\arabic{section}. #1}
}

\suppsection{Validation of the cross-stream migration model}
\label{sec-2}

\begin{figure}
  \centering{\includegraphics[width=1.0\columnwidth]{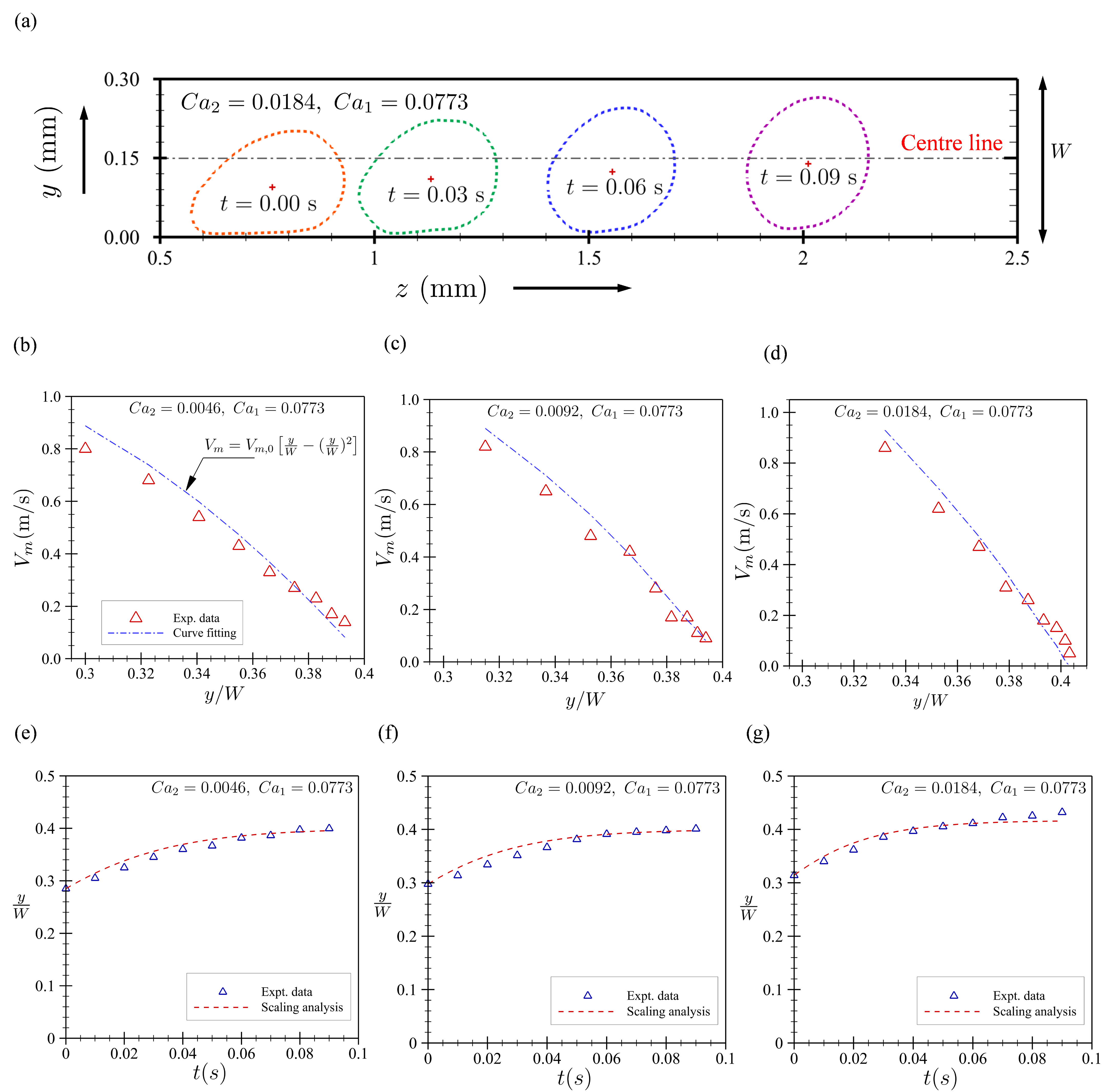}}
   \medskip
  \caption{\justifying{Validation of the predicted cross-stream migration of the primary droplet against experimental measurements. (a) Time sequence illustrating the evolution of droplet shape and its migration across the channel for a representative ($Ca_2 = 0.0184$, $Ca_1=0.0733$, and $Ec=10.87$) condition. (b–d) Cross-stream migration velocity of the primary droplet as a function of time for three different $Ca_2$ (= 0.0046, 0.0092, and 0.0184) at fixed $Ca_1$ (= 0.0773), showing comparison between model predictions and experiments. (e–g) Corresponding radial displacement of the primary droplet towards the channel centreline, comparing theoretical predictions with experimental measurements for the same flow conditions.}}
\label{fig:S1}
\end{figure}

To assess the predictive capability of the proposed model, we compare the theoretically predicted cross-stream migration dynamics of the primary droplet with experimental measurements over a range of flow conditions spanning the squeezing and dripping regimes. During the splitting process, the primary droplet experiences both a non-inertial lift force (NIL) \cite{Abkarian2002} and an elasticity-induced lift force \citep{Hazra2019}. The combined action of these forces generates a net transverse force that drives the droplet towards the channel centreline (see figure~\ref{fig:S1}(a)). The results presented in this section correspond to the breakup of a highly elastic fluid ($Ec=10.87$). Unless otherwise stated, the continuous-phase capillary number is fixed at $Ca_1=0.0773$, while the dispersed-phase capillary number, $Ca_2$, is varied over the range $0.0046$--$0.0184$.

In the main text (\S 3.2.2), we proposed a scaling for the initial migration velocity of the droplet near the wall as: $V_{m,0} \approx \left[ Ca^2 + (1/\beta)^{0.33}~Wi \right]\left(r_p/W\right)^2 u_{avg}$. As the droplet migrates across the channel width, the resultant lift force evolves due to changes in the droplet geometry and the local elastic stresses, thereby altering the migration velocity and radial position of the droplet. Consequently, predicting the droplet trajectory requires detailed knowledge of the time-dependent droplet shape and stress distribution. To avoid this complexity, we adopt a simplified approach and track the temporal evolution of the droplet centroid as it migrates towards the channel centreline in the $y$-direction under different flow conditions.

Figure~\ref{fig:S1}(a) illustrates the temporal evolution of the droplet shape and its migration towards the channel centreline (dash-dot line) for a representative case ($Ca_2=0.0184$, $Ca_1=0.0773$, and $Ec=10.87$). Here, the dotted contour represents the droplet interface, while the plus symbol (+) inside the droplet denotes its centroid, located at ($x, y$). As time progresses, the droplet is advected downstream and simultaneously migrates towards the channel centreline. The cross-stream migration velocity of the droplet is therefore determined from the instantaneous centroid positions measured experimentally as $V_m =\Delta y/\Delta t$. Figures~\ref{fig:S1}(b--d) present the variation of $V_m$ with the dimensionless radial position ($y/W$) for different values of $Ca_2$ while keeping $Ca_1$ fixed. To quantify the dependence of $V_m$ on $y/W$, the experimental data are fitted with an empirical relation, which yields $V_m = V_{m,0} [(y/W)-2.5(y/W)^2]$.

Having obtained the migration velocity profile, $V_m = V_{m,0} [(y/W)-2.5(y/W)^2]$, we predict the radial displacement of the droplet by numerically integrating the kinematic equation governing its cross-stream motion. Specifically, the droplet position is advanced in time using the Forward Euler method as $y_{i+1}=y_i+V_m(y_i)\Delta t$, where $y_i$ is the radial position of the droplet centroid at time $t_i$, $V_m(y_i)$ is the corresponding migration velocity obtained from the empirical fit, and $\Delta t$ is the time step. The integration is initiated from the experimentally measured initial centroid position, $y_0$, and is subsequently performed over the time interval $t=0$ to $t_n$. Since the initial migration velocity, $V_{m,0}$, is known from the proposed scaling relation and the variation of $V_m$ with $y/W$ is obtained from the experimental fitting, the droplet trajectory can be determined iteratively. This procedure yields the temporal evolution of the droplet radial position as it migrates towards the channel centreline. The predicted trajectory is then compared with the experimentally measured radial displacement to assess the validity of the proposed scaling framework.

For validation, the experimentally measured radial position, $y/W$, is plotted as a function of time, $t$, and compared with the predictions obtained from the proposed scaling relation combined with the numerical integration procedure under different flow conditions. As shown in figure~\ref{fig:2}(e--g), the predicted trajectories are in good agreement with the experimental measurements throughout the migration process. The ability of the model to accurately reproduce the temporal evolution of the droplet radial position demonstrates that the proposed scaling framework captures the dominant physical mechanisms governing the cross-stream migration of the primary droplet.

\begin{figure}
  \centering{\includegraphics[width=0.80\columnwidth]{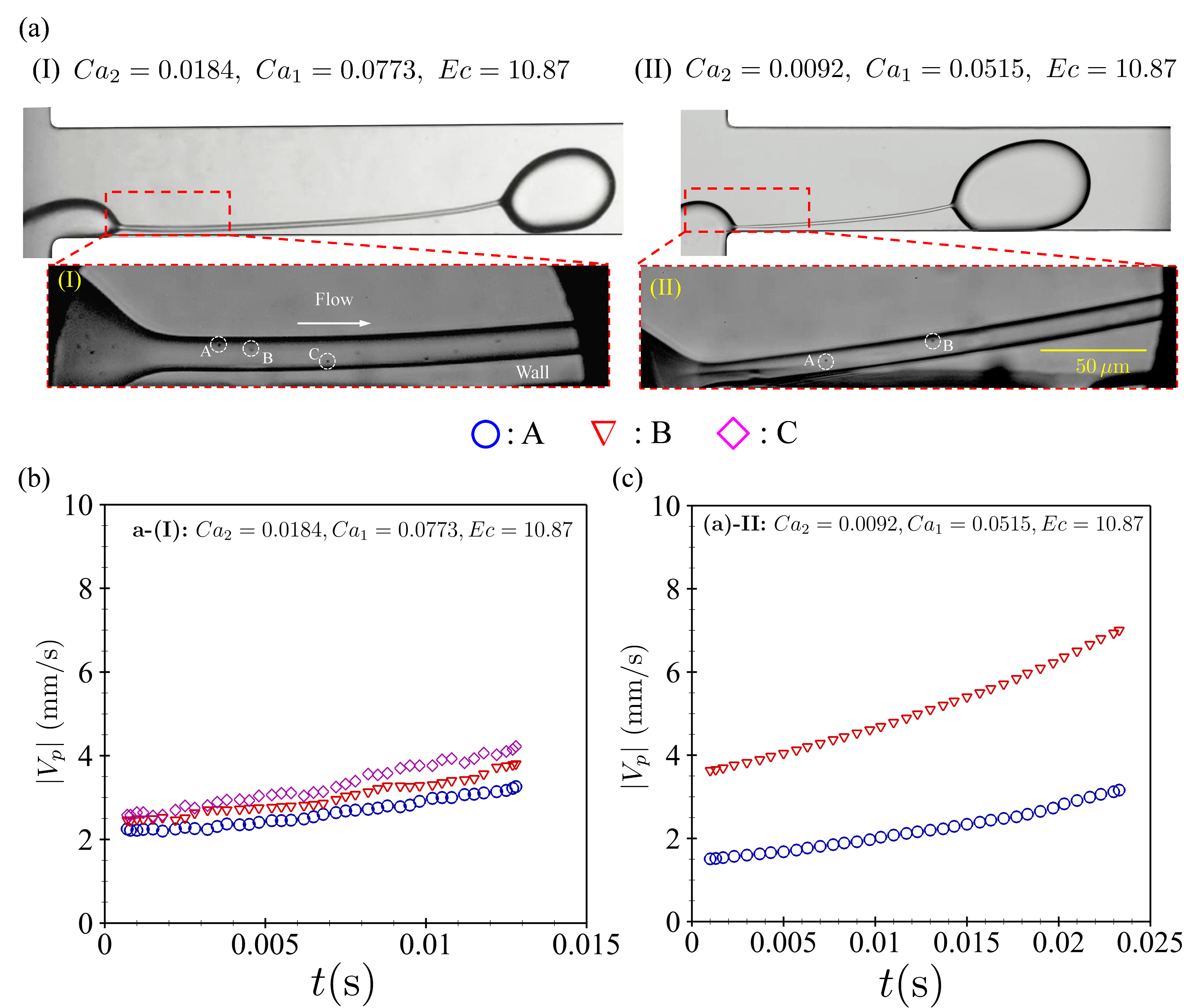}}
   \medskip
  \caption{\justifying{Influence of filament inclination on the internal velocity field of a highly elastic fluid ($Ec = 10.87$), determined from the trajectories of $1,\mu\text{m}$ tracer particles. (a) Representative experimental images of a stable filament under two flow conditions. (I) At $Ca_2 = 0.0184$ and $Ca_1 = 0.0773$, a portion of the filament remains nearly parallel to the channel wall near the junction; the magnified view (60×) identifies the tracer particles used for velocity measurements. (II) At $Ca_2 = 0.0092$ and $Ca_1 = 0.0515$, the filament exhibits a pronounced inclination, and the enlarged view shows tracer particles located at different distances from the wall. (b) Normalized velocity magnitude of particles A, B, and C positioned at approximately the same wall-normal location, showing only minor deviations relative to the reference velocity of particle B ($|V_P|_B$). (c) Velocity magnitude of a particle located farther from the wall (B) compared with that of a particle closer to the wall (A), highlighting the effect of the velocity gradient across the inclined filament, with the velocity increasing by approximately two times away from the wall.}}
\label{fig:S2}
\end{figure}

\suppsection{Additional evidence for the effect of filament inclination on the internal velocity field}
In \S 3.4 of the main text, we demonstrated that filament inclination generates a velocity gradient across the filament thickness, with fluid elements located farther from the channel wall moving at higher velocities than those closer to the wall. This velocity asymmetry was shown to play a key role in the elasticity-induced migration of the primary droplet towards the channel centreline.

To further illustrate this mechanism within the range of capillary numbers corresponding to the squeezing and dripping regimes identified in the regime map, we present additional measurements for a different combination of capillary numbers in Figure~\ref{fig:S2}. The local fluid velocity within the filament was quantified by tracking $1~\mu$m tracer particles. Consistent with the observations reported in the main text, particles located at approximately the same wall-normal position exhibit nearly matching velocities, whereas particles positioned at different distances from the wall experience markedly different velocities. In particular, particles located farther from the wall move significantly faster than those closer to the wall, indicating the presence of a strong velocity gradient across the inclined filament.

The qualitative agreement between the present results and those presented in \S 3.4 (main text) confirms that the velocity asymmetry induced by filament inclination is a robust feature of the breakup dynamics of highly elastic fluids. These additional measurements therefore provide further support for the physical mechanism proposed in the main text and reinforce the connection between filament inclination, asymmetric fluid transport, and the resulting elasticity-induced migration of the primary droplet.

\end{document}